\documentstyle[epsfig,12pt]{article}
\newcommand{\beq}{\begin{equation}}
\newcommand{\eeq}{\end{equation}}
\newcommand{\beqa}{\begin{eqnarray}}
\newcommand{\eeqa}{\end{eqnarray}}
\newcommand{\beqan}{\begin{eqnarray*}}
\newcommand{\eeqan}{\end{eqnarray*}}
\newcommand{\ba}{\begin{array}}
\newcommand{\ea}{\end{array}}
\newcommand{\ben}{\begin{enumerate}}
\newcommand{\een}{\end{enumerate}}
\newcommand{\bfl}{\begin{flushleft}}
\newcommand{\efl}{\end{flushleft}}
\newcommand{\btab}{\begin{tabular}}
\newcommand{\etab}{\end{tabular}}
\newcommand{\bit}{\begin{itemize}}
\newcommand{\eit}{\end{itemize}}
\newcommand{\bdes}{\begin{description}}
\newcommand{\edes}{\end{description}}
\newcommand{\bdm}{\begin{displaymath}}
\newcommand{\edm}{\end{displaymath}}
\newcommand{\nl}{\nonumber \\}
\newcommand{\no}{\nonumber}
\newcommand{\ul}{\underline}
\newcommand{\ol}{\overline}
\newcommand{\ra}{\rightarrow}

\newcommand{\ve}{\varepsilon}
\newcommand{\vp}{\varphi}

\newcommand{\dg}{\dagger}

\newcommand{\wh}{\widehat}

\newcommand{\Ha}{{\cal H}}

\newcommand{\cL}{{\cal L}}

\newcommand{\dfrac}{\displaystyle \frac}

\newcommand{\lets}{\stackrel{<}{_\sim}}
\newcommand{\del}{\partial}

\newcommand{\toG}{\stackrel{G}{\to}}

\normalsize
\begin{document}
\parskip=4pt plus 1pt  
\begin{titlepage}
\begin{flushright}
UWThPh-1996-34\\
June 1996
\end{flushright}
\vspace{2cm}
\begin{center}
{\Large \bf
CHIRAL PERTURBATION THEORY} \\[30pt]
{\bf Gerhard Ecker} \\[10pt]
Institut f\"ur Theoretische Physik, Universit\"at Wien \\[5pt]
Boltzmanngasse 5, A--1090 Wien, Austria \\[50pt]
{\bf ABSTRACT} \\[10pt]
\end{center}
\noindent
After a general introduction to the structure of effective field theories,
the main ingredients of chiral perturbation theory are reviewed. Applications
include the light quark mass ratios and pion--pion scattering to two--loop
accuracy. In the pion--nucleon system, the linear $\sigma$ model is
contrasted with chiral perturbation theory. The heavy--nucleon
expansion is used to construct the effective pion--nucleon
Lagrangian to third order in the low--energy expansion, with applications
to nucleon Compton scattering.

\vfill
\begin{center}
Lectures given at the \\[5pt]
V$^{\rm th}$ Workshop on Hadron Physics, Angra dos Reis, RJ, Brazil\\[5pt]
April 15 - 20, 1996 \\[5pt]
To appear in the Proceedings
\end{center}
\vfill


\end{titlepage}
\setcounter{page}{1}
\tableofcontents

\renewcommand{\thesection}{\arabic{section}}
\renewcommand{\thesubsection}{\arabic{section}.\arabic{subsection}}
\renewcommand{\theequation}{\arabic{section}.\arabic{equation}}

\setcounter{equation}{0}
\setcounter{subsection}{0}
 
\section{The Standard Model at Low Energies}
\label{sec:SM}

These lecture notes consist of three parts. In the first lecture, some basic
properties of chiral perturbation theory (CHPT) \cite{Wein79,GL84,GL85a,Leu94} 
are reviewed. After a general classification of effective field theories,
the nonlinear realization of spontaneously broken chiral symmetry
is discussed. Effective chiral Lagrangians and the loop expansion are 
described as the two main ingredients of a systematic low--energy
expansion. In the second lecture I discuss some applications in the 
purely mesonic sector. After a brief introduction to the phenomenology
at next--to--leading order, I concentrate
on our present knowledge of the light quark masses and on recent
calculations of elastic $\pi\pi$ scattering to $O(p^6)$, i.e. up to
two loops. In the last lecture, nucleons and pions are treated together. 
The linear $\sigma$ model is contrasted with CHPT. The heavy--nucleon 
expansion is introduced to construct the effective pion--nucleon Lagrangian 
to third order in the low--energy expansion. As an instructive application, 
some aspects of nucleon Compton scattering are considered at the end. 
More extended treatments of CHPT can be found in recent reviews 
\cite{BKM95,Eck95,Pich95,EdR95}.

\subsection{Effective field theories}
\label{subsec:EFT}

Effective field theories incorporate our intuitive knowledge that  
quantum gravity is irrelevant for understanding the pion--pion
scattering phase shifts. They are the down--to--earth alternative to the
Theory of Everything and they provide a theoretical framework for most of 
particle physics and for other fields as well. The notion of an effective 
theory comes with an energy scale $\Lambda$ that separates the regimes of 
effective and ``fundamental'' descriptions. Whenever we do not know the
underlying theory, the effective field theory approach can be used to
parametrize the unknown physics at smaller distances.  
But even if the ``fundamental'' theory is known
it is often unnecessary to consider all dynamical degrees of
freedom at the same footing. Instead, it may be more convenient for 
doing phenomenology at low energies to integrate out 
the heavy degrees of freedom that cannot be accessed directly at the energies
in question. In many cases, the structure at small distances can be
transferred to the effective level by the technique of ``matching''.
In other cases, this is not possible, usually because perturbation
theory breaks down at the transition energy. In fact, these
lectures deal with such a case: QCD in the confinement regime.

It is useful to distinguish two classes of effective field theories.
\subsection*{i. Decoupling effective field theories}
This is the standard case where nothing much is happening in
the transition from the fundamental to the effective level. As
the heavy degrees of freedom are integrated out, only the light degrees 
of freedom already present in the theory remain. In terms of the
light fields, the effective Lagrangian has the general form
\beq
\cL_{\rm eff} = \cL_{d\leq 4} + \sum_{d>4} \frac{1}{\Lambda^{d-4}}
\sum_{i_d} g_{i_d} O_{i_d}~.  \label{eq:EFT}
\eeq
The first part $\cL_{d\leq 4}$ contains the potentially renormalizable terms
with operator dimension $d \leq 4$: the terms with  $d < 4$ are called
relevant and those with $d = 4$ marginal operators. The second part
contains the so--called irrelevant operators with $d > 4$;
the $g_{i_d}$ are dimensionless coupling constants expected to be
at most of $O(1)$,
and the $O_{i_d}$ are monomials in the light fields with operator 
dimension $d$. At energies much below $\Lambda$,
corrections due to the nonrenormalizable parts ($d>4$) are suppressed
by powers of $E/\Lambda$.
However, the usual nomenclature is sometimes misleading: in many
cases of interest, some of which are listed below, it is precisely the
``irrelevant'' part of the effective Lagrangian that contains all
the interesting physics.

There are many examples of this type:
\bit
\item QED for $E \ll m_e$ \cite{EuHe36}.
\item The Fermi theory of weak interactions for $E \ll M_W$. In
both examples, all the interest is in the irrelevant parts of the effective
Lagrangian starting with $d=8$ in the Euler--Heisenberg Lagrangian and
$d=6$ for the Fermi theory.
\item The Standard Model itself is in all likelihood an effective field
theory. In this case, we neither know the scale $\Lambda$ nor have
we caught any glimpse of the interesting ``irrelevant'' part yet.
\item The scattering of light on neutral atoms for small photon energies
can also be treated with an effective field theory. I refer to 
Ref.~\cite{Kaplan95} for an explanation in field theory terms of why the 
sky is blue.
\eit

\subsection*{ii. Non--decoupling effective field theories}
In this second case, 
the transition from the fundamental to the effective level occurs through a
phase transition via the spontaneous breakdown of a symmetry generating
Goldstone bosons that usually acquire a small mass
($M \ll \Lambda$) due to some explicit symmetry breaking (pseudo--Goldstone
bosons). A spontaneously broken symmetry relates processes with different 
numbers of Goldstone bosons. Such a symmetry
transformation is therefore nonlinear in the Goldstone fields and 
the distinction between renormalizable ($d\leq 4$) and nonrenormalizable
($d>4$) parts in the effective Lagrangian (\ref{eq:EFT}) would
not be preserved under a symmetry transformation.
There is no intrinsic difference between relevant, marginal and
irrelevant operators in this case.
Although we shall encounter a well--known exception
in the form of the linear $\sigma$ model \cite{SGML57},
the effective Lagrangian in the non--decoupling case is
generically nonrenormalizable. 

The Goldstone theorem \cite{Gold61} does not only predict the existence
of massless excitations, but it also requires that the interactions of
Goldstone bosons vanish as their energies tend
to zero. This is the basis for a systematic low--energy expansion of 
effective Lagrangians of type ii: instead of the operator dimension as
in (\ref{eq:EFT}), the number of derivatives (powers of momenta in 
momentum space) distinguishes successive terms in the Lagrangian.

The general structure of effective Lagrangians with spontaneously broken
symmetries is largely independent of the specific physical
realization. Two well--known examples in particle physics
are the Standard Model without an explicit Higgs boson (heavy--Higgs
scenario) where a gauge symmetry is spontaneously broken and 
QCD at energies below 1 GeV where the global chiral symmetry is 
spontaneously broken. The universality of Goldstone
boson interactions implies that the scattering of longitudinal gauge vector 
bosons is in first approximation analogous to $\pi\pi$ scattering. 
The comparison between these two examples may carry an additional message.
The Higgs sector of the Standard Model is modelled after the linear
$\sigma$ model of low--energy hadron physics. As I will explain in
the third lecture, we now know that the linear $\sigma$ model is not 
the effective field theory for QCD. The realistic effective field theory 
at low energies (CHPT) can do without an explicit scalar
field. Could this be a lesson for the scalar sector at the Fermi scale?

\subsection{Chiral symmetry}
\label{subsec:CHS}

QCD with $N_f$ massless quarks exhibits a global symmetry
$$
\underbrace{SU(N_f)_L \times SU(N_f)_R}_{\mbox{chiral group $G$}}
\times U(1)_V \times U(1)_A ~.
$$
At the effective hadronic level, the quark
number symmetry $U(1)_V$ is realized as baryon number. The axial 
$U(1)_A$ is not a symmetry at the quantum level due to the  Abelian
anomaly \cite{HCC76}.

A classical symmetry can be realized in quantum field theory
in two different ways depending on how the vacuum responds
to a symmetry transformation (Wigner--Weyl vs. Nambu--Goldstone).
There is a large body of theoretical and phenomenological evidence that the
chiral group $G$ is spontaneously broken to the vectorial subgroup
$H=SU(N_f)_V$ [isospin for $N_f=2$, flavour $SU(3)$ for $N_f=3$]. 
The axial generators of $G$ are nonlinearly realized
and there are $N_f^2 - 1$ massless pseudoscalar Goldstone bosons
to be identified with the low--lying pseudoscalar mesons.

Spontaneous symmetry breaking $G\to H$ can be characterized by one
or more order parameters. In quantum field theory, they take the form of
vacuum expectation values
\beq
\langle 0|A|0\rangle \label{eq:op} 
\eeq
of operators $A$ that are invariant under the conserved subgroup
$H$ but transform non--trivially under the full group $G$. If the vev
(\ref{eq:op}) is non--vanishing, the vacuum cannot be invariant
under $G$ signalling spontaneous
symmetry breaking. Which are the possible order parameters for chiral
symmetry breaking in QCD? The operators $A$ must be
Lorentz scalar (even parity), colour singlet, $G$ non--invariant, but
$SU(N_f)_V$ invariant quark--gluon operators. Ordering the candidate
operators according to their dimension, one finds a unique possibility
with the smallest possible dimension $d=3$: $A= {\overline q}q$
where $q=u,d (,s)$ for $N_f=2 (3)$. The order parameters are called
quark condensates:
\beq
\langle 0|{\overline u}u|0\rangle =\langle 0|{\overline d}d|0\rangle (=
\langle 0|{\overline s}s|0\rangle) \ne 0 ~.
\label{eq:cond}
\eeq
The equalities in (\ref{eq:cond}) are a consequence of $SU(N_f)_V$ invariance.
Increasing the dimension, there is again a unique operator with $d=4$,
the square of the gluon field strength that is however a $G$ singlet
and therefore not relevant for an order parameter. Also for $d=5$,
there is a single candidate ${\overline q}\lambda^a \sigma_{\mu\nu}
G_a^{\mu\nu}q$ giving rise to the so--called mixed condensate.
For $d \ge 6$, there are many more possibilities such as four--quark operators.

The original formulation of CHPT \cite{GL84,GL85a} assumes that the quark
condensates have the values extracted from many different applications
of QCD sum rules \cite{QCDSR} but also from lattice simulations \cite{GB96},
giving rise to the dominant contributions for the masses of the pseudoscalar
mesons. These lectures deal with the standard formulation of CHPT.
In the not very likely case that all the sum rule and lattice evidence
is misleading and the quark condensates are
much smaller or even zero, an alternative framework called Generalized
CHPT \cite{GCHPT} might become more appropriate. Occasionally, I
will come back to this option.

There is a standard procedure how to implement a symmetry
transformation on the Goldstone fields \cite{CCWZ69}.
Geometrically, the Goldstone fields $\vp$ can be viewed as coordinates of
the coset space $G/H$. An element $g$ of the symmetry group $G$ induces in a
natural way (by left translation) a transformation of $u(\vp)\in G/H$:
\beq 
u(\vp) \toG g u(\vp) = u(\vp') h(g,\vp)~.  \label{eq:coset}
\eeq
The so--called compensator field $h(g,\vp)$ is an element of the conserved
subgroup $H$ and it accounts for the fact that a coset element is
only defined up to an $H$ transformation. For $g \in H$, the
symmetry is realized in the usual linear way (Wigner--Weyl) and $h(g)$
does not depend on the Goldstone fields $\vp$. On the other hand,
for $g \in G$ corresponding to a spontaneously broken symmetry 
($g \not\in H$), the symmetry is realized nonlinearly (Nambu--Goldstone)
and $h(g,\vp)$ does depend on $\vp$.

For the special case of chiral symmetry $G=SU(N_f)_L \times SU(N_f)_R$,
parity relates left-- and right--chiral transformations. With a standard
choice of coset representatives the general transformation 
(\ref{eq:coset}) then takes the special form
\beq
 u(\vp')=g_R u(\vp) h(g,\vp)^{-1} 
= h(g,\vp) u(\vp) g_L^{-1} \label{eq:uphi}
\eeq
$$
g=(g_L,g_R) \in G~.
$$
Since the chiral coset space $SU(N_f)_L \times SU(N_f)_R/SU(N_f)_V$,
even though it is not a group, is homeomorphic to $SU(N_f)$ as a manifold,
$u(\vp) \in SU(N_f)$ is a matrix--valued field.
Different forms (the most familiar one being the exponential parametrization)
correspond to different coordinate systems. A field transformation amounts
to a coordinate transformation in coset space.

For practical purposes, one never needs to know the explicit form
of $h(g,\vp)$, but only the transformation property
(\ref{eq:uphi}). In the mesonic sector, it is often more convenient
to work with the square of $u(\vp)$. Because of (\ref{eq:uphi}), the 
matrix field $U(\vp)=u(\vp)^2$ has a simpler linear transformation behaviour:
\beq
U(\vp) \toG g_R U(\vp) g_L^{-1} \label{eq:Uphi}~.
\eeq
It is therefore frequently used as basic building block for chiral
Lagrangians. When non--Goldstone degrees of freedom like baryons or
meson resonances are included in the effective Lagrangians, the 
nonlinear picture with $u(\vp)$
and $h(g,\vp)$ is more appropriate. In the third part of these 
lectures, the abstract quantities introduced here will emerge 
naturally when we investigate the linear $\sigma$ model in some detail.

Before embarking on the construction of effective chiral Lagrangians,
we recall that there is actually no chiral symmetry in nature. In addition
to the spontaneous breaking, chiral symmetry is explicitly broken
both by non--vanishing quark masses and by the electroweak interactions.
Both conceptually and for practical purposes, the best way to keep track
of the explicit breaking is through the introduction of external matrix
fields \cite{GL84,GL85a} $v,a,s,p$. The QCD Lagrangian $\cL^0_{\rm QCD}$ for
massless quarks is extended to
\beq
\cL = \cL^0_{\rm QCD} + \bar q \gamma^\mu(v_\mu + a_\mu \gamma_5)q -
\bar q (s - ip \gamma_5)q \label{eq:QCD}
\eeq
to include electroweak interactions of quarks with external gauge fields
$v,a$ and to allow for nonzero quark masses by setting the scalar matrix 
field $s(x)$ equal to the diagonal quark mass matrix. 
One performs all calculations with a (locally) $G$ invariant effective
Lagrangian in a manifestly chiral invariant manner. Only at the very 
end, one inserts the appropriate
external fields to extract the Green functions of quark
currents or matrix elements of interest. The explicit breaking of
chiral symmetry is automatically taken care of by this spurion technique.
In addition, electromagnetic gauge invariance is manifest. All
Ward identities for Green functions of quark currents are guaranteed.

Although this procedure produces all Green functions for electromagnetic
and weak currents, the method must be extended in order to include
virtual photons (electromagnetic corrections) or virtual $W$ bosons
(nonleptonic weak interactions). 
 
\subsection{Chiral Lagrangians}
\label{subsec:CHL}

CHPT is based on a two--fold expansion. As a
low--energy effective field theory, it is an expansion in small
momenta. On the other hand, it is also an expansion in quark masses 
around the chiral limit. In full generality,
the effective chiral Lagrangian is of the form
\beq
\cL_{\rm eff} = \sum_{i,j} \cL_{ij}~, \qquad \qquad
\cL_{ij} = O(p^i m^j_q)~. \label{eq:gexp}
\eeq
The two expansions become related by expressing the pseudoscalar meson
masses in terms of the quark masses $m_q$. If the quark condensate is
non--vanishing in the chiral limit, the squares of the meson masses
start out linear in $m_q$ [cf. Eq.~(\ref{eq:mpi2})].
Assuming the linear terms to give the dominant contributions to the
meson masses, one arrives at the standard chiral counting \cite{GL84,GL85a}
with $m_q = O(p^2)$  and
\beq
\cL_{\rm eff} = \sum_d \cL_d~, \qquad \qquad
\cL_d = \sum_{i + 2j = d} \cL_{ij}~.\label{eq:Leff}
\eeq

As already mentioned, there are many indications in favour of the standard
picture. In addition to QCD sum rules and lattice simulations,
the Gell-Mann--Okubo mass formula (\ref{eq:GMO}) for the pseudoscalar 
meson masses can only be understood in a natural way if the squares of
the meson masses are linear in the quark masses to a good approximation.
In order to account for the logical possibility that the 
quark condensate would not give the main contributions to the meson masses,
the proponents of Generalized CHPT \cite{GCHPT} suggest a
reordering of the effective Lagrangian. Through this reordering, more 
terms appear at a given order that are relegated to higher orders in the 
standard counting. Therefore, more unknown constants appear
at any given order compared to the standard framework.
Nevertheless, the effective chiral Lagrangian is 
the same in both approaches. In these lectures, I will adhere to the 
standard procedure, but I will refer once more to Generalized CHPT when
discussing $\pi\pi$ scattering at the two--loop level.

The construction of the effective chiral Lagrangian for the strong
interactions of mesons is now straightforward. In terms of the basic building
blocks $U(\vp)$ and the external fields $r_\mu=v_\mu+a_\mu$, 
$l_\mu=v_\mu-a_\mu$, $s$ and $p$ and with the standard chiral counting
just described, the chiral invariant Lagrangian starts
out at $O(p^2)$ with
\beq
\cL_2 = \frac{F^2}{4} \langle D_\mu U D^\mu U^\dg + \chi U^\dg +
\chi^\dg U \rangle \label{eq:L2}
\eeq
$$ 
\chi = 2B(s + ip) \qquad \qquad 
D_\mu U = \partial_\mu U - i  r_\mu U + i U l_\mu 
$$
where $\langle \dots \rangle$ stands for the $N_f-$dimensional trace.
The two low--energy constants (LECs) at $O(p^2)$ are related to the pion 
decay constant and to the quark condensate in the chiral limit:
\beq
F_\pi =  F[1 + O(m_q)] = 92.4 ~{\rm MeV} \label{eq:FB}
\eeq
$$\langle 0|\bar u u |0\rangle = - F^2 B[1 + O(m_q)] ~.
$$
Expanding the Lagrangian (\ref{eq:L2}) to second order in the meson
fields and setting the external scalar field equal to the quark mass 
matrix, one can immediately read off the pseudoscalar meson masses
to leading order in $m_q$, e.g.,
\beq
M^2_{\pi^+}  = (m_u + m_d) B~.\label{eq:mpi2}
\eeq
As expected, the squares of the meson masses are linear in the
quark masses to leading order if the quark condensate is non--vanishing in the
chiral limit ($B \neq 0$). The full set of equations for the masses
of the pseudoscalar octet gives rise to several well--known relations:
\beqa
F_\pi^2 M_\pi^2 = - (m_u + m_d) \langle 0|\bar u u|0\rangle &
\mbox{      } & \cite{GMOR68} \label{eq:GMOR} \\
B = \frac{M_\pi^2}{m_u+m_d} = \frac{M_{K^+}^2}{m_s + m_u} =
\frac{M_{K^0}^2}{m_s + m_d} & \mbox{      } & \cite{Wein77} 
\label{eq:ratios} \\
3M^2_{\eta_8} = 4 M_K^2 - M_\pi^2 & \mbox{      } & \cite{GMO57}. 
\label{eq:GMO}
\eeqa

The effective chiral Lagrangian of the Standard Model is shown 
schematically in Table \ref{tab:EFTSM}. The subscripts of the different
parts of this Lagrangian denote the chiral dimension according to
the standard counting and the numbers in brackets indicate the appropriate
number of LECs. The notation even/odd refers to the
mesonic Lagrangians without/with an $\varepsilon$ tensor (even/odd intrinsic
parity). For instance, ${\cal L}_4^{\rm odd}(0)$ stands for the anomalous
Lagrangian of the
Wess--Zumino--Witten functional \cite{WZW71} that has no free parameters.
I have grouped together those pieces of the Lagrangian that
have the same chiral order as a corresponding loop amplitude ($L$ =
0, 1, 2). For a given $L$, the first line contains purely mesonic Lagrangians
and the second one refers to the pion--nucleon system.
The Lagrangians ${\cal L}_n^{\Delta S=1}$ and ${\cal L}_n^\gamma$ 
describe nonleptonic weak interactions and virtual photons, respectively.
There are similar Lagrangians for the meson--baryon system which I have
not included in the Table. In the meson--baryon sector only
the pion--nucleon Lagrangian is included, i.e. $N_f=2$. On the other
hand, the numbers of LECs in the purely mesonic Lagrangians are given 
for $N_f=3$. The theory has to be renormalized for $L\ge 1$.
The parts of the effective chiral Lagrangian that
have been completely renormalized are underlined in Table \ref{tab:EFTSM}.

\renewcommand{\arraystretch}{1.1}
\begin{table}[t]
\begin{center}
\caption{The effective chiral Lagrangian of the Standard Model}
\label{tab:EFTSM}
\vspace{.5cm}
\begin{tabular}{|lc|c|} \hline
\hspace{2cm} ${\cal L}_{\rm chiral \,\, dimension}$ ~($\#$ of LECs)  
& \hspace{.2cm} & loop order \\[10pt] 
\hline 
& & \\[10pt]
${\cal L}_2(2)$~+~${\cal L}_4^{\rm odd}(0)$~+~${\cal L}_2^{\Delta S=1}(2)$
~+~${\cal L}_0^\gamma(1)$  & \hspace{.2cm} & $L=0$ \\[10pt]
~+~${\cal L}_1^{\pi N}(1)$~+~${\cal L}_2^{\pi N}(7)$~+~\dots & & \\[20pt]
~+~$\ul{{\cal L}_4^{\rm even}(10)}$~+~$\ul{{\cal L}_6^{\rm odd}(32)}$
~+~$\ul{{\cal L}_4^{\Delta S=1}(22,{\rm octet})}$~+
~$\ul{{\cal L}_2^\gamma(14)}$ & \hspace{.3cm} &  $L=1$ \\[10pt]
~+~$\ul{{\cal L}_3^{\pi N}(24)}$~+~${\cal L}_4^{\pi N}(?)$~+~\dots & & \\[20pt]
~+~${\cal L}_6^{\rm even}(111)$~+~\dots & & $L=2$ \\[12pt] \hline
\end{tabular}
\end{center}
\end{table}

Why do we have to add special Lagrangians
for the nonleptonic weak interactions and for virtual photons?
After all, the strong chiral Lagrangian like (\ref{eq:L2}) contains
external photons and $W$ bosons. However, unlike for semileptonic
processes where we can hook on a leptonic current (electromagnetic or
weak) to the external gauge fields, the strong interactions cannot be
disentangled from the electroweak interactions in the
nonleptonic case. Put in another way, Green functions of quark currents are not 
sufficient to generate nonleptonic weak amplitudes or amplitudes
with virtual photons.

Let us first consider the nonleptonic weak interactions.
At the low energies relevant for CHPT, the correct procedure is
to first integrate out the $W$ together with the heavy quarks to arrive
at an effective Hamiltonian already at the quark level \cite{SVZ77,GW79}:
\beq
\Ha_{\rm eff}^{\Delta S =1} = \frac{G_F}{\sqrt{2}} V_{ud} V_{us}^*
\sum_i C_i(\mu) Q_i + {\rm h.c.} \label{eq:Hnl}
\eeq
The $C_i(\mu)$ are Wilson coefficients depending on the QCD 
renormalization scale $\mu$. The $Q_i$ are local four--quark operators
if we limit the operator product expansion (\ref{eq:Hnl}) to the
leading $d = 6$ operators. For the effective realization at the
hadronic level, the explicit form of the $Q_i$ (or of the Wilson
coefficients) is of no concern. All that is needed is the transformation
property of $\Ha_{\rm eff}^{\Delta S = 1}$ under chiral rotations:
\beq
\Ha_{\rm eff}^{\Delta S =1} \sim (8_L,1_R) + (27_L,1_R)~.
\label{eq:nldecomp}
\eeq
The task is then to construct the most general chiral Lagrangian
that has the same transformation property (\ref{eq:nldecomp}) under
chiral transformations. As indicated in Table \ref{tab:EFTSM}, the 
lowest--order Lagrangian is again of $O(p^2)$ and it has two LECs,
one for the octet piece and one for the 27--plet.

Integrating out the photons cannot be described
by a local operator at low energies as in the case of the massive
$W$ boson. In a first step, the electromagnetic
field is made dynamical by including the appropriate kinetic term and by
enlarging the external vector field $v_\mu$:
\beq
v_\mu \ra v_\mu - e Q A_\mu~, \label{eq:dyn}
\eeq
where $A_\mu$ is the dynamical photon field and the quark charge matrix
$Q$ is given in (\ref{eq:Qquark}).
CHPT then generates automatically all diagrams with virtual (and real)
photons. However, this is not the whole story. For instance, loop
diagrams with virtual photons will in general be divergent requiring 
appropriate local counterterms. If we restrict our attention to single--photon
exchange, we must add the most general chiral Lagrangian of
$O(e^2)$ that transforms as the product of two electromagnetic currents
under chiral rotations. For $N_f = 3$, the electromagnetic current is
pure octet. The easiest method for constructing the appropriate chiral 
Lagrangian is based on the so--called spurion technique \footnote{The same
procedure can be employed for the nonleptonic weak Lagrangian.}. One  
writes down the most general chiral invariant Lagrangian that is bilinear in 
octet spurion fields $Q_L(x),Q_R(x)$ with transformation properties
\beq
Q_A(x) \toG g_A Q_A(x) g^{-1}_A, \qquad A = L,R~.
\eeq
Identifying the spurion fields with the quark charge matrix,
\beq
Q_A(x) = Q = \mbox{ diag}(2/3,-1/3,-1/3)~,\label{eq:Qquark}
\eeq
gives rise to the effective chiral Lagrangian of $O(e^2)$ with the correct
transformation properties. The lowest--order Lagrangian ${\cal L}_0^\gamma(1)$
is of $O(p^0)$ in this case and it has a single coupling constant.
We will come back to this Lagrangian in the discussion of $\pi\pi$
scattering in the second lecture.

\subsection{Loop expansion}
\label{subsec:loops}
Now that we know how to construct effective chiral Lagrangians we can
calculate tree--level amplitudes to any desired order in the
low--energy expansion. Although in the sixties many proponents of effective 
Lagrangians argued that due to their nonrenormalizability such Lagrangians 
only make sense at tree level, it is clear that tree--level amplitudes
cannot be consistent with sacred principles of
quantum field theory like unitarity and analyticity (except at
lowest order in the chiral expansion). For instance, unitarity requires
for the forward elastic meson--meson scattering amplitude $T$
\beq
\Im m ~T \ge |T|^2~.
\eeq
Since the (real) amplitude starts at $O(p^2)$, there must be an imaginary
part at $O(p^4)$. This imaginary part cannot come from a hermitian
Lagrangian at tree level, but can only be due to loop diagrams.
Thus, the loop expansion is essential for a consistent low--energy expansion.

Today, we view effective field theories on almost the same footing as 
``fundamental" gauge theories (see also Ref.~\cite{GW95}). They admit a 
perfectly well--defined loop expansion although they are nonrenormalizable.
The nonrenormalizability manifests itself in the appearance of
additional terms that are not present in the lowest--order Lagrangian.
However, $\cL_{\rm eff}$ is already the most general chiral 
Lagrangian with the appropriate transformation properties. 
Since the divergences can be absorbed by local counterterms
that exhibit the same symmetries as the initial Lagrangian, $\cL_{\rm eff}$
automatically includes all terms needed for renormalization to
every order in the loop expansion. 

Before we can renormalize the theory, we have to regularize it. In principle,
any regularization that respects chiral symmetry is good enough.
In practice, a mass independent
regularization scheme like dimensional regularization is best
suited for the purpose. In addition to respecting chiral symmetry,
it very much simplifies the chiral counting 
and it avoids spurious quadratic or higher divergences.

\begin{figure}
\centerline{\epsfig{file=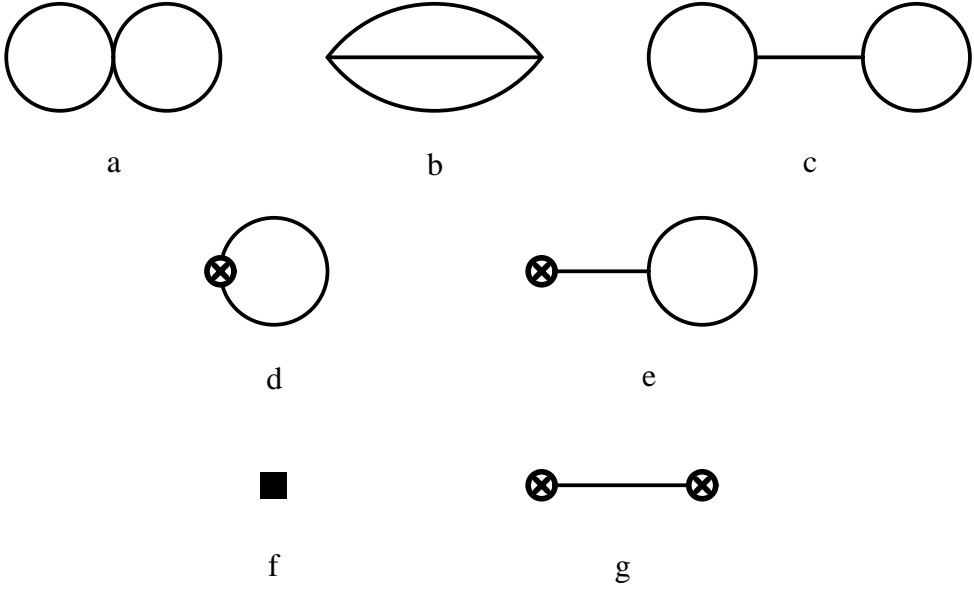,height=8cm}}
\caption{Skeleton diagrams of $O(p^6)$. Normal vertices are from
$\cL_2$, crossed circles denote vertices from $\cL_4$ and the square in diagram
f stands for a vertex from $\cL_6$.
The propagators and vertices carry the full tree structure associated
with the lowest--order Lagrangian $\cL_2$.}
\label{fig:p6}
\end{figure}

To keep track of the chiral counting, it is convenient to
define the chiral dimension of an amplitude. In the mesonic case
with only pseudoscalar mesons in
internal lines, the chiral dimension $D$ of a connected $L$--loop
 amplitude with $N_d$ vertices of $O(p^d)$ ($d = 2,4,\ldots)$ is given by
\cite{Wein79}
\beq
D = 2L + 2 + \sum_d (d-2) N_d~, \qquad d = 4,6,\ldots \label{eq:DL}
\eeq
As this formula shows, the number $N_2$ of vertices from the lowest--order
Lagrangian (\ref{eq:L2}) does not affect the chiral dimension. 
The following classification of amplitudes with chiral dimension up to 
$O(p^6)$ is therefore independent of $N_2$:
\beqa
D = 2 : & L = 0  & \no\\*
D = 4 : & L = 0, & N_4 = 1 \no\\* 
        & L =1 & \no\\*
D = 6 : & L = 0, & N_6 = 1 \no\\*
        & L = 0, & N_4 = 2 \nl
	& L = 1, & N_4 = 1 \nl
	& L = 2 & ~. \label{eq:D246}
\eeqa
The basic topological structures for $D=6$ are shown in Fig.~\ref{fig:p6}.
In any of these diagrams arbitrary tree structures from the lowest--order
Lagrangian (\ref{eq:L2}) can emerge from both the lines and the vertices.
For instance, to find all diagrams contributing to $\pi\pi$ scattering
at $O(p^6)$ one hooks on four external pion lines to every skeleton
diagram in Fig.~\ref{fig:p6} in all possible allowed ways. We come
back to $\pi\pi$ scattering in the second lecture.

For a given amplitude, the chiral dimension $D$ increases with
$L$ according to Eq.~(\ref{eq:DL}). In order to reproduce the (fixed)
physical dimension of the amplitude, each loop produces a factor $1/F^2$.
Together with the geometric loop factor $(4\pi)^{-2}$, the loop expansion
suggests
\beq
4\pi F_\pi = 1.2 \mbox{ GeV} \label{eq:naive}
\eeq
as the natural scale of the chiral expansion \cite{MG84}. A more refined 
analysis leads to
\beq
4\pi F_\pi/\sqrt{N_f}\label{eq:soph}
\eeq
as the relevant scale for $N_f$ light flavours \cite{SS90}. Restricting
the domain of applicability of CHPT to momenta $|p| \lets 
O(M_K)$, the natural expansion parameter of chiral amplitudes
based on the naive estimate (\ref{eq:naive}) is
expected to be of the order
\beq
\frac{M_K^2}{16 \pi^2 F_\pi^2} = 0.18~. \label{eq:omag}
\eeq
In addition, these terms often appear multiplied with chiral logarithms. It 
is therefore no surprise that substantial higher--order
 corrections  in the chiral expansion are the rule rather than the exception
for chiral $SU(3)$. On the other hand, for 
$SU(2)_L\times SU(2)_R$ and for momenta 
$|p| \lets O(M_\pi)$ the chiral expansion is expected to converge
considerably faster.

It is rather obvious that chiral invariant Lagrangians together
with a chiral invariant regularization and renormalization procedure
will produce amplitudes that respect all chiral Ward identities. It is 
expected, but highly non--trivial nevertheless, that the converse is also 
true. As shown by Leutwyler \cite{Leu94}, the most general solution
of chiral Ward identities can always be generated by a locally chiral
invariant $\cL_{\rm eff}$. The only exception is the chiral anomaly that 
requires a Wess--Zumino--Witten term \cite{WZW71}, but the rest 
is gauge invariant (see also Ref.~\cite{DHW94}).

\setcounter{equation}{0}
\setcounter{subsection}{0}

\section{The Physics of Mesons}
\label{sec:meson}

\subsection{Phenomenology at next--to--leading order}
\label{subsec:p4}
The Green functions and amplitudes of lowest order $p^2$ (current
algebra level) are determined by the Lagrangian (\ref{eq:L2}) at
tree level. At next--to--leading order, $O(p^4)$, there are three
types of contributions \cite{GL84,GL85a} in accordance with (\ref{eq:D246}):
\bit
\item Tree diagrams with a single vertex from the effective chiral 
Lagrangian $\cL_4$ of $O(p^4)$ given below and, as usual, any number of 
vertices from $\cL_2$.
\item One--loop diagrams with all vertices from $\cL_2$.
\item The Wess--Zumino--Witten functional \cite{WZW71}
to account for the chiral ano\-ma\-ly \cite{ABJB69}.
\eit
 
For $N_f=3$, the effective chiral Lagrangian $\cL_4(U,v,a,s,p)$ 
has the form \cite{GL85a}
\beqa
{\cal L}_4 & = & L_1 \langle D_\mu U^\dg D^\mu U\rangle^2 +
                 L_2 \langle D_\mu U^\dg D_\nu U\rangle
                     \langle D^\mu U^\dg D^\nu U\rangle \no \\*
& & + L_3 \langle D_\mu U^\dg D^\mu U D_\nu U^\dg D^\nu U\rangle +
    L_4 \langle D_\mu U^\dg D^\mu U\rangle \langle \chi^\dg U +
    \chi U^\dg\rangle  \no \\*
& & +L_5 \langle D_\mu U^\dg D^\mu U(\chi^\dg U + U^\dg
\chi)\rangle
    +
    L_6 \langle \chi^\dg U + \chi U^\dg \rangle^2 +
    L_7 \langle \chi^\dg U - \chi U^\dg \rangle^2  \no \\*
& & + L_8 \langle \chi^\dg U \chi^\dg U +
 \chi U^\dg \chi U^\dg\rangle
    -i L_9 \langle F_R^{\mu\nu} D_\mu U D_\nu U^\dg +
      F_L^{\mu\nu} D_\mu U^\dg D_\nu U \rangle \no \\*
& & + L_{10} \langle U^\dg F_R^{\mu\nu} U F_{L\mu\nu}\rangle +
    L_{11} \langle F_{R\mu\nu} F_R^{\mu\nu} + F_{L\mu\nu} F_L^{\mu\nu}\rangle +
    L_{12} \langle \chi^\dg \chi \rangle
\label{eq:L4}
\eeqa
where $F_R^{\mu\nu}$, $F_L^{\mu\nu}$ are the field strength tensors
associated with the external gauge fields $r_\mu$, $l_\mu$.
This is the most general Lorentz invariant Lagrangian of $O(p^4)$ with
local chiral symmetry, parity and charge conjugation. 

Before discussing the new LECs $L_1, \ldots, L_{12}$,
let us consider one--loop diagrams. Since the theory is nonrenormalizable
there are divergences all over the place. General results
of renormalization theory and the chiral counting given in (\ref{eq:D246})
ensure that those divergences can be absorbed in a local Lagrangian
of $O(p^4)$ that has all the symmetries of the lowest--order Lagrangian
(\ref{eq:L2}). But the Lagrangian (\ref{eq:L4}) is the most general
such Lagrangian. Therefore, all one--loop divergences can be
absorbed by divergent LECs $L_i$ in 
(\ref{eq:L4}). The specific values depend of course on the regularization 
procedure. Remember also that we are using a mass independent regularization 
scheme. With a mass dependent regularization, the lowest--order LECs $F$
and $B$ would in general receive quadratically divergent one--loop
contributions. Since physically relevant quantities must be independent
of the regularization scheme, those quadratic divergences are completely
spurious effects because they do not arise in dimensional
regularization for instance.

Subtracting the divergent parts of the $L_i$ introduces a scale dependence
of the measurable LECs $L_i^r(\mu)$. This scale dependence is always
compensated by the analogous scale dependence of one--loop diagrams.
The latter appears in the form of so--called chiral logs
$\sim \ln{p^2/\mu^2}$. The corresponding scale
dependence of the LECs takes the form
\beq
L_i^r(\mu_2) = L_i^r(\mu_1) + \frac{\Gamma_i}{(4\pi)^2} \ln
\frac{\mu_1}{\mu_2} \label{eq:scale}~.
\eeq
The coefficients $\Gamma_i$, the $\beta$--functions of the $L_i$, are
listed in Table \ref{tab:Li}.
 
\renewcommand{\arraystretch}{1.1}
\begin{table}[t]
\begin{center}
\caption{Phenomenological values and source for the renormalized coupling
constants $L^r_i(M_\rho)$, taken from Ref.~\protect\cite{BEG95}.
The quantities $\Gamma_i$
in the fourth column determine the scale dependence of the $L^r_i(\mu)$
according to Eq.~(\protect\ref{eq:scale}). $L_{11}^r$ and $L_{12}^r$ are not
directly accessible to experiment.} \label{tab:Li}
\vspace{.5cm}
\begin{tabular}{|c||r|l|r|}  \hline
i & $L^r_i(M_\rho) \times 10^3$ & source & $\Gamma_i$ \\ \hline
  1  & 0.4 $\pm$ 0.3 & $K_{e4},\pi\pi\rightarrow\pi\pi$ & 3/32  \\
  2  & 1.35 $\pm$ 0.3 &  $K_{e4},\pi\pi\rightarrow\pi\pi$&  3/16  \\
  3  & $-$3.5 $\pm$ 1.1 &$K_{e4},\pi\pi\rightarrow\pi\pi$&  0     \\
  4  & $-$0.3 $\pm$ 0.5 & Zweig rule &  1/8  \\
  5  & 1.4 $\pm$ 0.5  & $F_K:F_\pi$ & 3/8  \\
  6  & $-$0.2 $\pm$ 0.3 & Zweig rule &  11/144  \\
  7  & $-$0.4 $\pm$ 0.2 &Gell-Mann--Okubo,$L_5,L_8$ & 0             \\
  8  & 0.9 $\pm$ 0.3 & \small{$M_{K^0}-M_{K^+},L_5,$}&
5/48 \\
     &               &   \small{ $(2m_s-m_u-m_d):(m_d-m_u)$}       & \\
 9  & 6.9 $\pm$ 0.7 & $\langle r^2\rangle^\pi_V$ & 1/4  \\
 10  & $-$5.5 $\pm$ 0.7& $\pi \rightarrow e \nu\gamma$  &  $-$ 1/4  \\
\hline
11   &               &                                & $-$1/8 \\
12   &               &                                & 5/24 \\
\hline
\end{tabular}
\end{center}
\end{table}
 
The renormalized coupling constants $L_i^r(\mu)$ are measurable quantities
that characterize QCD at $O(p^4)$. The present values of the $L_i^r(\mu)$
extracted from phenomenology are displayed in Table \ref{tab:Li}. Once
these values are established, one can make predictions for all Green
functions and amplitudes of $O(p^4)$ in terms of the $L_i^r(\mu)$ 
(actually only the LECs with $i=1,\dots,10$ appear in physical quantities).
This is not the place to discuss in detail the phenomenology at $O(p^4)$ for
which I refer to Refs.~\cite{BKM95,Eck95,Pich95,EdR95}. By and
large, the theoretical work has been completed at next--to--leading
order even though experimental results are not yet available in
all cases. 

For reactions involving an odd number of mesons, there are no contributions
at $O(p^2)$. The leading order is $p^4$ and it is unambiguously
given by the Wess--Zumino--Witten functional \cite{WZW71}. The 
next--to--leading order, $O(p^6)$ in this case, involves again one--loop
contributions and a Lagrangian $\cL_6^{\rm odd}$. As indicated in Table
\ref{tab:EFTSM}, this Lagrangian has 32 new LECs. Unlike in the 
even--intrinsic--parity
sector, most of those LECs are not yet known. One can make estimates
based on meson resonance exchange, which works very well
for the LECs $L_i$ of $O(p^4)$ \cite{EGPR89}. For a review of the ``anomalous'' sector
at next--to--leading order, I refer to Ref.~\cite{Bij93a}.

Let me finish this part with a remark on the structure of Green functions
and amplitudes of $O(p^4)$. Obviously, the
chiral dimension $D=4$ does not imply that the relevant Green functions
are just fourth--order polynomials in external momenta and masses. Instead,
the chiral dimension has to do with the degree of homogeneity of
the amplitudes in momenta and masses. Consider the Feynman amplitude 
$A$ for a general process with $D_F$ external photons and
$W$ bosons (semileptonic transitions). If we define in addition
to (\ref{eq:DL})
\beq
D = D_L + D_F  ~,
\label{eq:DF}
\eeq
then $D_L$ is the degree of homogeneity of the amplitude $A$ as 
a function of external momenta ($p$) and meson masses ($M$):
\beq
A(p,M;C_i^r(\mu),\mu/M) = M^{D_L} \, A ( p/M , 1;C_i^r(\mu), \mu/M )  ~.
\label{eq:homog}
\eeq
The $C_i^r(\mu)$ denote renormalized LECs. In the meson sector at 
$O(p^4)$, they are just the $L_i^r(\mu)$, but the structure (\ref{eq:homog})
is completely general.

\subsection{Light quark masses}
\label{subsec:mq}
Quark masses are fundamental parameters of QCD. The current
quark masses of the QCD Lagrangian depend on the QCD renormalization 
scale. Since the effective chiral Lagrangians cannot depend on
this scale, the quark masses always appear multiplied by quantities
that transform contragrediently under changes of the renormalization
scale. The chiral Lagrangian (\ref{eq:Leff}) contains the quark masses
via the scalar field $\chi$ defined in (\ref{eq:L2}). As long as one
does not use direct or indirect information on $B$ [related to the
quark condensates, cf. Eqs.~(\ref{eq:cond}), (\ref{eq:FB})], one can 
only extract ratios of quark masses.

\begin{figure}
\centerline{\epsfig{file=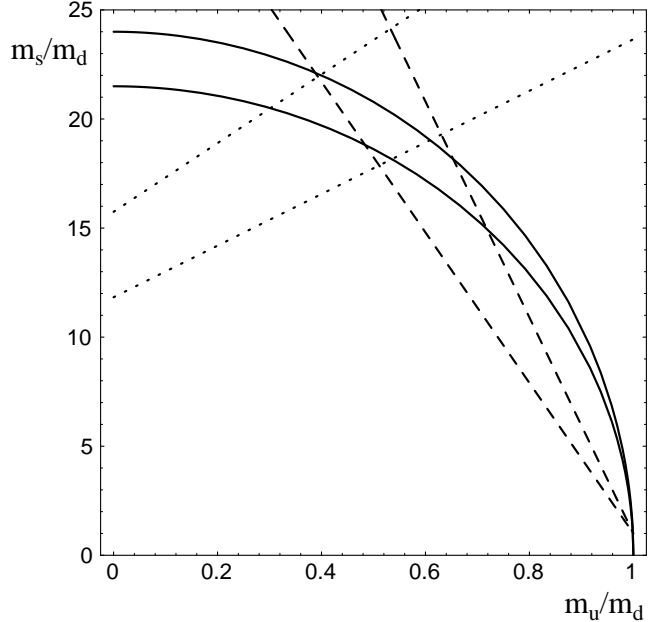,height=8cm}}
\caption{First quadrant of Leutwyler's ellipse \protect\cite{Leu90}
for $Q=24$ (upper curve) and $Q=21.5$ (lower curve). The dotted lines
correspond to $\Theta_{\eta \eta'}=-15^0$ (upper line) and $-25^0$
(lower line). The wedge between the two dashed lines corresponds
to $35 \leq R \leq 50$ with $R$ defined in (\protect\ref{eq:Rq}).}
\label{fig:ellipse}
\end{figure}

The lowest--order mass formulas together with Dashen's
theorem on the lowest--order electromagnetic contributions to the
meson masses \cite{Dash69} lead to the ratios \cite{Wein77}
\beq
\frac{m_u}{m_d} = 0.55~, \qquad \qquad
\frac{m_s}{m_d} = 20.1 ~. \label{eq:mratio}
\eeq
These ratios receive higher--order corrections. The most important
ones are corrections of $O(p^4) = O(m_q^2)$ and $O(e^2 m_s)$. The
corrections of $O(p^4)$ were worked out by Gasser and Leutwyler \cite{GL85a}
who found that the ratios ($2 \hat{m}:= m_u+m_d$)
\beqa
\frac{M_K^2}{M_\pi^2} &=& \frac{m_s + \hat{m}}{m_u + m_d} [1 + \Delta_M +
O(m_s^2)] \label{eq:MKpi} \\
\frac{(M_{K^0}^2 - M_{K^+}^2)_{\rm QCD}}{M_K^2 - M_\pi^2} &=&
\frac{m_d - m_u}{m_s - \hat{m}} [1 + \Delta_M + O(m_s^2)]
\eeqa
depend on the same correction $\Delta_M$ of $O(m_s)$:
\beq
\Delta_M = \frac{8(M_K^2 - M_\pi^2)}{F^2} (2 L_8^r - L_5^r) +
\mbox{ chiral logs~.} \label{eq:DelM}
\eeq
Since the ratio of these two ratios is independent of $\Delta_M$, one
can express the quantity
\beq
Q^2 := \frac{m_s^2 - \hat{m}^2}{m_d^2 - m_u^2} \label{eq:Qratio}
\eeq
in terms of masses only, at least up to higher--order corrections:
\beq
Q^2 = \frac{M_K^2}{M_\pi^2} \cdot
\frac{M_K^2 - M_\pi^2}{M_{K^0}^2 - M_{K^+}^2 + M_{\pi^+}^2 - M_{\pi^0}^2}
\cdot \left[1 + O(m_s^2) + O\left( e^2 \frac{m_s}{m_d - m_u}
\right) \right].
\eeq
Without the corrections, this relation implies 
\beq
Q = Q_D = 24.2 ~.
\eeq
Plotting $m_s/m_d$ versus $m_u/m_d$ leads to
Leutwyler's ellipse \cite{Leu90} which is to a very good
approximation given by
\beq
\frac{1}{Q^2} \left( \frac{m_s}{m_d}\right)^2 + \left(
\frac{m_u}{m_d}\right)^2 = 1~.
\eeq
In Fig.~\ref{fig:ellipse}, the relevant quadrant of the ellipse is shown
for $Q = 24$ (upper curve) and $Q = 21.5$ (lower curve).

However, there are possibly large corrections of $O(e^2 m_s)$. Assuming
iso\-spin symmetry ($m_u=m_d$), the quantity
\beq
\Delta_{\rm EM} = M^2_{K^+} - M^2_{K^0} - M^2_{\pi^+} + M^2_{\pi^0}
\label{eq:DEM}
\eeq
vanishes to lowest order, $O(e^2 p^0)$, according to Dashen's theorem
\cite{Dash69}. The one--loop calculation was performed independently by 
Urech \cite{Ur95} and by Neu\-feld and Rupertsberger \cite{NR95}:
\beqa
\Delta_{\rm EM} &=& - e^2 M^2_K \left[ \frac{1}{(4\pi)^2} \left( 3 \ln
\frac{M^2_K}{\mu^2} - 4 + \frac{2C}{F^4} \ln \frac{M^2_K}{\mu^2}\right)
 + \frac{16C}{F^4} L_5^r(\mu) + S_\Delta^r(\mu)\right] \no \\
&& \mbox{} + O(e^2 M^2_\pi)~.
\eeqa
Here, $C$ is the single LEC of $O(e^2 p^0)$ [cf. ${\cal L}_0^\gamma(1)$
in Table \ref{tab:EFTSM}] and $S_\Delta^r(\mu)$ is a combination of
LECs of $O(e^2 p^2)$ corresponding to  
${\cal L}_2^\gamma(14)$ in Table \ref{tab:EFTSM}. Since $S_\Delta^r(\mu)$
is unknown, one can only calculate the scale dependent remainder without 
additional assumptions. The conclusion is that $\Delta_{\rm EM}$ may be 
sizable but the strong scale dependence precludes a more definite answer.

To get a better feeling for the size of $\Delta_{\rm EM}$, one
must turn to more model dependent approaches. In Table \ref{tab:DEM}
the existing predictions of different models are displayed.
In summary, the earlier calculations found a rather big value while
the more recent ones tend to give smaller values. The model of
Ref.~\cite{DHW93} was criticized by Baur and Urech \cite{BU96} to
violate chiral symmetry. The provisional overall conclusion at
this time is that $\Delta_{\rm EM}$ is probably positive implying
$Q < Q_D$ by at most 10$\%$.

\renewcommand{\arraystretch}{1.1}
\begin{table}[t]
\begin{center}
\caption{Model calculations for the quantity $\Delta_{\rm EM}$ defined
in (\protect\ref{eq:DEM}).} \label{tab:DEM}
\vspace{.5cm}
\begin{tabular}{|c|c|c|}  \hline
\mbox{   } $\Delta_{\rm EM}\cdot 10^3$ GeV$^{-2}$ \mbox{   } & \mbox{   } 
main ingredient \mbox{   } & \mbox{   } Ref. \mbox{   } \\ \hline
  1.2  & $V,A$ exchange & \protect\cite{DHW93}  \\
  1.3 $\pm$ 0.4  & large $N_c$ & \protect\cite{Bij93b}  \\
  0.6 & lattice & \protect\cite{DET96}     \\
  $-$ 0.13 $\div$ 0.36  & $V,A$ exchange & \protect\cite{BU96}  \\
\hline
\end{tabular}
\end{center}
\end{table}

There is completely independent information on $Q$ from $\eta \to 3
\pi$ decays. In contrast to the masses, electromagnetic corrections
are known to be tiny in this case \cite{BS66,BKW96}. In terms of $Q$,
there is a parameter free prediction for the rate to $O(p^4)$ \cite{GL85c}:
\beq
\Gamma (\eta \to \pi^+ \pi^- \pi^0) = \Gamma_0 \left(\dfrac{Q_D}{Q}\right)^4
\eeq
with
$$
\Gamma_0 = (168 \pm 50)~\mbox{eV}~.
$$
The error in $\Gamma_0$ includes an estimate of the final state interactions.
Those final state interactions have recently been calculated
by two groups using dispersion relations. Both find a
substantial increase of $\Gamma_0$:
\beq
\Gamma_0 = \left\{\ba{ll}
(209 \pm 20)~\mbox{eV}& \qquad \cite{KWW96} \\
(219 \pm 22)~\mbox{eV}& \qquad \cite{AL96}
\ea \right.~.
\eeq
Using the present experimental value $\Gamma(\eta \to \pi^+\pi^-\pi^0)=
(283 \pm 28)$ eV \cite{PDG94}, one obtains 
\beq
Q = \left\{\ba{ll}
22.4 \pm 0.9 & \qquad \cite{KWW96} \\
22.7 \pm 0.8 & \qquad \cite{AL96}
\ea \right.
\eeq
confirming the positive value of $\Delta_{\rm EM}$ in Table \ref{tab:DEM}.

The ellipse in Fig.~\ref{fig:ellipse} is therefore well established.
What about the separate mass ratios $m_u/m_d$ and $m_s/m_d$ ? As emphasized
by Kaplan and Manohar \cite{KM86}, the ratios
cannot be determined separately from low--energy data alone due to an
accidental symmetry of $\cL_2 + \cL_4$. The
chiral Lagrangian is invariant to this order under the transformations
\\[5pt]
\centerline{
$m'_u = \alpha_1 m_u + \alpha_2 m_d m_s$ \qquad
 (and cyclic permutations)}
\beq
B' = B/\alpha_1, \qquad L'_6 = L_6 - \alpha, \qquad 
L'_7 = L_7 - \alpha, \qquad L'_8 = L_8 + 2\alpha \label{eq:KM}
\eeq
$$
\alpha = \frac{\alpha_2 F^2}{32 \alpha_1 B}~. 
$$
The corresponding terms in $\cL_2 + \cL_4$ involve only the scalar
field $\chi$, but no external gauge fields.
Consequently, all Green functions of vector and axial--vector currents
and thus all S--matrix elements are invariant under the
transformations (\ref{eq:KM}). Since $\Delta_M$ in
(\ref{eq:DelM}) is not invariant, it cannot be determined from
$V,A$ ~Green functions and S--matrix elements alone. It is important 
to realize that the symmetry (\ref{eq:KM}) is certainly not a (hidden) 
symmetry of QCD. For instance, the matrix element
\beq
\langle 0|{\overline d}i \gamma_5 u|\pi^+\rangle =\sqrt{2}F_\pi
\dfrac{M_{\pi^+}^2}{m_u+m_d}\label{eq:pseudo}
\eeq
is not invariant under this symmetry although it is a perfectly
well--defined quantity in QCD \cite{Leu96b}. Unfortunately, there
are no direct experimental probes for (pseudo)scalar densities as for 
(axial--)vector currents, but the matrix element (\ref{eq:pseudo}) is
at least in principle calculable, for instance on the lattice.

Some further input is therefore necessary to get an estimate for the
quantity $\Delta_M$. One possibility is to appeal to the successful
notion of resonance saturation \cite{EGPR89}, in this case for
the two--point functions of pseudoscalar quark densities \cite{Leu90}. 
Expressing $L_5$ and $L_8$
through the well--measured corrections to $F_K/F_\pi$ and to the
Gell-Mann--Okubo formula (\ref{eq:GMO}), $\Delta_M$ can be written 
in the form 
\beq
\Delta_M = - \frac{32 (M_K^2 - M_\pi^2)}{F^2} \; L_7 - 0.33~,
\label{eq:DMnum}
\eeq
reducing the problem to a determination of the scale independent 
LEC $L_7$.
Various arguments lead to a consistent picture for $L_7$ corresponding
to the value given in Table \ref{tab:Li}. 
$L_7$ also accounts for $\eta - \eta'$ mixing at $O(p^4)$ \cite{GL85a}.
A given value of $\Theta_{\eta \eta'}$ defines a straight line in $m_s/m_d$ 
vs. $m_u/m_d$. The dotted
lines in Fig.~\ref{fig:ellipse} correspond to $\Theta_{\eta \eta'} =
- 15^0$ (upper line) and $- 25^0$ (lower line), respectively. Note that
$\Delta_M =0$ corresponds to $\Theta_{\eta \eta'} = - 22^0 $ in perfect
agreement with the experimentally observed value. Although
the destructive interference between the two terms in (\ref{eq:DMnum})
enhances the uncertainty, the conclusion is that $\Delta_M$ is indeed
small as expected for a higher--order correction.

Another possibility to get a handle on $\Delta_M$ is
to make use of the large--$N_c$ expansion because the Kaplan--Manohar
transformation (\ref{eq:KM}) violates the Zweig rule. In other words,
results to a given order in $1/N_c$ are not invariant under the 
transformation. Working to first non--leading order in both CHPT and
$1/N_c$, Leutwyler \cite{Leu96a} has recently derived a lower bound
\beq
\Delta_M \ge - 0.07~.
\eeq
With very mild additional assumptions, he finds that $\Delta_M$
must actually be positive. As usual, one must take this large--$N_c$ 
result with a grain of salt. The complete expression for $\Delta_M$
in (\ref{eq:DelM}) is given by
\beq
\Delta_M = \frac{8(M_K^2 - M_\pi^2)}{F^2} [2 L_8^r(\mu) - L_5^r(\mu)] +
\dfrac{M_\eta^2 \ln(M_\eta^2/\mu^2)- M_\pi^2 \ln(M_\pi^2/\mu^2)}
{32 \pi^2 F^2}~.  \label{eq:DelMexp}
\eeq
The first part is leading in $N_c$ whereas the second part due to the
loop contributions is down by one power of $N_c$. For illustration,
let us take the mean values of the $L_i^r(M_\rho)$ as given in Table
\ref{tab:Li} and compare the two terms at $\mu=M_\rho$. The leading
first term comes out to be 0.09, but the subleading second term is
$- 0.05$. Thus, not only are the two terms comparable in size, but
they also interfere destructively. 

Altogether, a rather conservative bound
\beq
|\Delta_M| < 0.15 \label{eq:DMbound}
\eeq
emerges corresponding approximately to the domain in Fig.~\ref{fig:ellipse}
between the two dotted lines.
 
As also shown in Fig.~\ref{fig:ellipse}, the resulting values for the
quark mass ratios are completely consistent with independent information 
on the ratio
\beq
R = \frac{m_s - \hat{m}}{m_d - m_u} \label{eq:Rq}
\eeq
measuring the relative size of $SU(3)$ and isospin breaking. The information
comes from mass splittings in the baryon octet, $\rho - \omega$ mixing
and from the branching ratio $\Gamma(\psi' \ra \psi \pi^0)/\Gamma(\psi'
\ra \psi \eta)$. A recent update of the resulting range for $R$
can be found in Ref.~\cite{Leu96b}.

The main conclusion is that the allowed values for the quark mass ratios are 
still close to the Weinberg values (\ref{eq:mratio}). In particular,
the theoretically interesting possibility $m_u = 0$ is strongly excluded.
It would require $\Delta_M = -0.45$, in blatant contradiction to the
generous bound (\ref{eq:DMbound}). There is still no better summary
of the situation than given in Ref.~\cite{Leu90}~: \mbox{`` \dots
$m_u=0$} is an interesting way not to understand this world -- it is
not the only one.''

For the absolute magnitude of quark masses, the most reliable estimates 
are based on QCD sum rules.
The most recent determinations give for the running $\ol{\rm MS}$ masses
at a scale of 1 GeV 
\beq
m_u + m_d = (12 \pm 2.5)~\mbox{MeV}\qquad \cite{BPR95}
\eeq
and
\beq
m_s = \left\{\ba{ll}
(189 \pm 32)~\mbox{MeV}& \qquad \cite{JM95} \\
(171 \pm 15)~\mbox{MeV}& \qquad \cite{CDPS95}
\ea \right.~.
\eeq
A recent analysis of all available lattice data on the light quark masses
finds somewhat smaller values \cite{GB96}.

\subsection{$\pi\pi$ scattering to $O(p^6)$}
\label{subsec:pipi}
Pion--pion scattering is a fundamental process for CHPT
that involves only the pseudo--Goldstone bosons of chiral
$SU(2)$. In the limit of isospin conservation ($m_u=m_d$), 
the scattering amplitude for
\beq
\pi^a(p_a) + \pi^b(p_b) \to \pi^c(p_c) + \pi^d(p_d)
\eeq
is determined by a single scalar function $A(s,t,u)$ defined by the
isospin decomposition
\beqa
T_{ab,cd} &=& \delta_{ab}\delta_{cd} A(s,t,u) + \delta_{ac}\delta_{bd}
A(t,s,u) + \delta_{ad}\delta_{bc} A(u,t,s) \nl
A(s,t,u) &=& A(s,u,t) 
\eeqa
in terms of the usual Mandelstam variables $s,t,u$. The amplitudes 
$T^I(s,t)$ of definite isospin $(I = 0,1,2)$ in
the $s$--channel are decomposed into partial waves ($\theta$ is the
center--of--mass scattering angle):
\beq
T^I(s,t)=32\pi\sum_{l=0}^{\infty}(2l+1)P_l(\cos{\theta})t_l^I (s)~.
\eeq
Unitarity implies that in the elastic region $4M_\pi^2 \leq s
\leq 16M_\pi^2$ the partial--wave amplitudes $t_l^I$ can be described 
by real phase shifts $\delta_l^I$.
The behaviour of the partial waves  near threshold is of the form
\beq
\Re e\;t_l^I(s)=q^{2l}\{a_l^I +q^2 b_l^I +O(q^4)\}~,
\eeq
with $q$ the center--of--mass momentum.
The quantities $a_l^I$ and $b_l^I$ are called 
scattering lengths and slope parameters, respectively.

At lowest order in the chiral expansion, $O(p^2)$, the scattering amplitude
is given by the current algebra result \cite{Wein66}
\beq
A_2(s,t,u)=\dfrac{s-M_\pi^2}{F_\pi^2}~, \label{eq:A2}
\eeq
leading in particular to an $I=0$ S--wave scattering length
\beq
a_0^0=\dfrac{7 M_{\pi^+}^2}{32 \pi F_\pi^2} = 0.16~.\label{eq:a00}
\eeq
As discussed in the first lecture in the context of the loop expansion,
the chiral expansion for $SU(2)$ is expected to converge rapidly 
near threshold. On the other hand,
CHPT produces singularities in the quark mass expansion (chiral logs).
For an $L$--loop amplitude, the chiral logarithms appear
in a general amplitude as $(\ln{p^2/\mu^2})^n$ with $n\le L$.
Here, $p$ is a generic momentum and the dependence on the arbitrary
scale $\mu$ is as always compensated by the scale dependence of the 
appropriate LECs in the amplitude. 

The scattering amplitude of $O(p^4)$ \cite{GL83,GL84}
has the general structure
\beqa
F_\pi^4 A_4(s,t,u) =& c_1 M_\pi^4 + c_2 M_\pi^2 s + c_3 s^2 +
c_4 (t-u)^2 \nl
&+ F_1(s) + G_1(s,t) + G_1(s,u)~.
\eeqa
The functions $F_1$, $G_1$ are one--loop functions and the coefficients
$c_1,\dots,c_4$ of the crossing symmetric polynomial of $O(p^4)$
are given in terms of the appropriate LECs [called $l_i$ in the $SU(2)$ 
case] and of $\ln{M_\pi^2/\mu^2}$. 
Since four different LECs appear at this order, it is not surprising that
chiral symmetry does not put any further constraints on the $c_i$. With the
phenomenological values of the LECs $l_i^r(\mu)$, Gasser and Leutwyler
\cite{GL83,GL84} obtained $a_0^0=0.20 \pm 0.01$ to be compared with the
value $a_0^0=0.26 \pm 0.05$ extracted from experiment \cite{Nagel79}.
Thus, there is a sizable correction of $O(p^4)$
which can be attributed to a large extent to chiral logs.

The value of $a_0^0$ has some bearing on the mechanism of spontaneous
chiral symmetry breaking \cite{Stern93}. If the pseudoscalar meson masses 
are not dominated by the lowest--order contributions proportional to the quark
condensate [cf. Eq.~(\ref{eq:mpi2})], the standard quark mass ratios 
could be modified. This is precisely the option that Generalized CHPT proposes
to keep in mind. In the generalized approach, the scattering lengths 
cannot be absolutely predicted at $O(p^2)$, but they
depend in addition on the quark mass ratio $r=2 m_s/(m_u+m_d)$
\cite{Stern93}. Taking the experimental mean value for $a_0^0$ at 
face value would decrease the quark mass ratio $r$ from its generally 
accepted value 26 [cf. Eq.~(\ref{eq:mratio})] to about 10. 

After a period of little activity on the experimental
side, there are now good prospects for more precise
data in the near future. On the one hand, the forthcoming data
from $K_{e4}$ decays at the $\Phi$--factory DA$\Phi$NE in Frascati 
are expected to reduce the experimental uncertainty of $a_0^0$ to some
$5\%$ in one year of running \cite{BF95}. In addition, the recently
approved experiment DIRAC at CERN \cite{Adeva94} measuring the lifetime
of $\pi^+\pi^-$ atoms will determine the difference $|a_0^0 - a_0^2|$
of $S$--wave scattering lengths.
These experimental prospects have prompted two groups to calculate the
scattering amplitude of $O(p^6)$ \cite{Knecht95,BCEGS96}.
Knecht et al. \cite{Knecht95} have used dispersion relations to calculate the
analytically non--trivial part of the amplitude and they have fixed
four of the six subtraction constants using $\pi\pi$ scattering data at
higher energies. This approach does not yield the chiral logs which,
although contributing only to the polynomial
part of the amplitude, make an important numerical contribution.
The standard field theory calculation of CHPT \cite{BCEGS96} produces all 
those terms, but one has to calculate
quite a few diagrams with $L=0,1$ and 2 loops as shown in 
Fig.~\ref{fig:p6} to obtain the final amplitude.
The two groups agree completely on the absorptive parts
which are contained in the functions $F_2$,
$G_2$ in the general decomposition
\beqa
F_\pi^6 A_6(s,t,u) =& d_1 M_\pi^6 + d_2 M_\pi^4 s + d_3 M_\pi^2 s^2 +
d_4 M_\pi^2 (t-u)^2 + d_5 s^3 + d_6 s (t-u)^2 \no \\*
&+ F_2(s) + G_2(s,t) + G_2(s,u)~.
\eeqa
The coefficients $d_1,\dots,d_6$ of the general crossing symmetric
polynomial of $O(p^6)$ depend on the LECs $l_i^r(\mu)$ ($i=1,
\dots,4$) of $O(p^4)$, on the chiral logs $(\ln{M_\pi^2/\mu^2})^n ~(n=1,2)$
and on six combinations of the LECs of $O(p^6)$ corresponding to
the Lagrangian ${\cal L}_6^{\rm even}(111)$ \cite{FS96} in Table
\ref{tab:EFTSM}. Again, chiral symmetry does not constrain those
six combinations. However, both chiral dimensional analysis and saturation
by resonance exchange suggest values for these LECs
that do not affect the threshold parameters in a dramatic way, especially
not for the $S$--waves. The coefficients $d_i$ are dominated
on the one hand by the LECs $l_i^r(\mu)$ of $O(p^4)$ and by the chiral
logs. Eventually, the $O(p^6)$ amplitude together with precise data
will allow for a much improved determination of the $l_i^r(\mu)$.
In the dispersion theory approach \cite{Knecht95}, four
combinations of the $d_i$ are determined from sum rules using data
at higher energies. The results are then presented in terms of the two
remaining parameters one of which is especially sensitive to the
quark condensate.

\begin{figure}[t]
\unitlength1cm
\begin{picture}(2,1) \end{picture}
\epsfysize=8cm
\epsffile{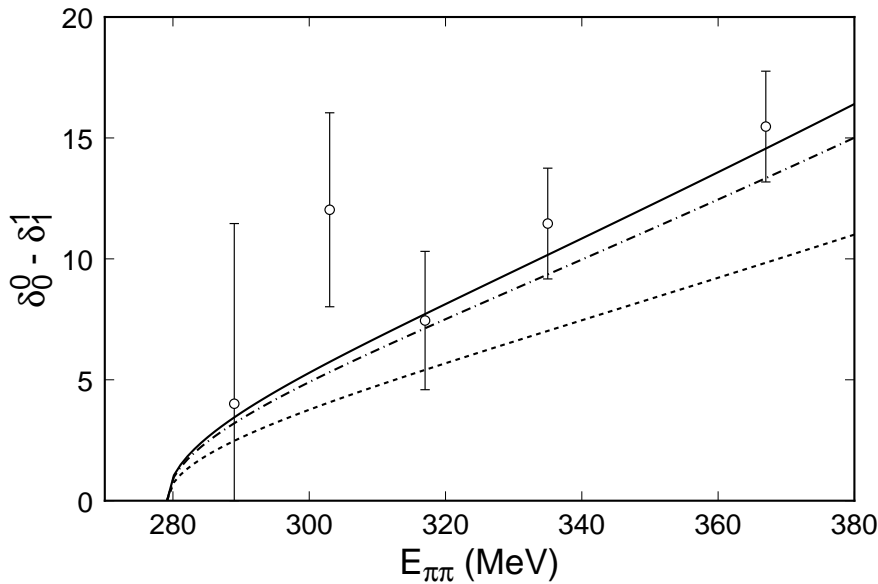}
\caption{The phase shift difference $\delta_0^0-\delta_1^1$ (in degrees) as a
function of the center of mass energy of the two incoming pions. The dotted
(dash--dotted) line displays the tree (one--loop) result,
whereas the solid line denotes the two--loop approximation 
\protect\cite{BCEGS96}. The data are from Ref.
\protect\cite{Ross77}.} \label{fig:phases}
\end{figure}

With the canonical values of the $l_i^r(\mu)$ \cite{GL84,BCG94}, 
I compare in Table \ref{tab:ab} the predicted values for the $S$--wave
threshold parameters for different chiral orders (using 
the ``old" value $F_\pi=93.2 ~{\rm MeV}$ and the charged pion mass).
In Fig.~\ref{fig:phases}, the phase shift difference $\delta_0^0 - \delta_1^1$
is plotted as a function of the center--of--mass energy of the
incoming pions.

\renewcommand{\arraystretch}{1.1}
\begin{table}[t]
\begin{center}
\caption{$S$--wave threshold parameters for elastic $\pi\pi$
scattering \protect\cite{Wein66,GL83,BCEGS96} in units of $M_{\pi^+}$. 
The experimental
values are taken from Ref.~\protect\cite{Nagel79}.}\label{tab:ab}

\vspace*{.5cm}

\begin{tabular}{|c|c|c|c|c|}
\hline 
 & & & &  \\
  &\mbox{  } $O(p^2)$\mbox{  } & \mbox{  }$O(p^4)$\mbox{  } 
&\mbox{  } $O(p^6)$ \mbox{  }&\mbox{  } experiment\mbox{  } \\[14pt] 
\hline 
 & & & & \\
$ a_0^0$& $0.16$ & $0.20$  & $0.22$  & $0.26\pm0.05$\\[10pt]

$b_0^0$ & $0.18$ & $0.25$& $0.275$  & $0.25\pm0.03$\\[10pt]

$-$10 $a_0^2$ & $0.45$ & $0.42$ & $0.41$  & $0.28\pm0.12$\\[10pt]

$-$10 $b_0^2$ & $0.91$ & $0.73$ & $0.72$ & $0.82\pm0.08$\\[10pt]
\hline
\end{tabular}
\end{center}
\end{table}

The main conclusion is that the corrections of
$O(p^6)$ are reasonably small and under theoretical control. For the
specific case of $a_0^0$, we observe that a value as high as $0.26$ 
is practically impossible to accommodate within standard CHPT. Therefore, 
the forthcoming data on $\pi\pi$ scattering will either corroborate
this prediction with significant precision or they will shed serious
doubts on the assumed mechanism of spontaneous chiral symmetry breaking 
through the quark condensate. 

At the level of precision we have reached with the $O(p^6)$ calculation,
one may wonder about the size of electromagnetic and isospin violating 
corrections. Neglecting the tiny $\pi^0-\eta$ mixing angle, the charged
and neutral pion masses are equal to $O(p^2)$ (without assuming
$m_u=m_d$):
\beq
M^2_{\pi^+}=M^2_{\pi^0}=(m_u+m_d) B ~.
\eeq
The mass difference
between charged and neutral pions is almost entirely an effect of
$O(e^2 p^0)$ and it is determined by the single LEC of the Lagrangian 
${\cal L}_0^\gamma(1)$
in Table \ref{tab:EFTSM}. This effect is certainly non--negligible also 
for $\pi\pi$
scattering as can be seen, for instance, from the lowest--order expression
for $a_0^0$ evaluated with either the
charged or the neutral pion mass:
\beq
a^0_0 = \dfrac{7 M^2_\pi}{32 \pi F^2_\pi} = \left\{\ba{ll}
0.156  &~\mbox{with} ~~M^2_{\pi^+} \\
0.146  &~\mbox{with} ~~M^2_{\pi^0}
\ea \right.~. \label{eq:a00+0}
\eeq
Clearly, the difference is comparable to the chiral correction of $O(p^6)$.
The pion decay constant is also affected by radiative corrections, which
have been estimated \cite{Holstein90,PDG94}
to move $F_\pi$ down from 93.2 to 92.4 MeV. This decrease of $F_\pi$
increases the $O(p^6)$ prediction for $a_0^0$ from 0.217 to 0.222.

The question is then what other effects of $O(e^2 p^0)$ appear in
the $\pi\pi$ scattering amplitude. To answer this question, we expand
the Lagrangian ${\cal L}_0^\gamma(1)$ in pion fields, setting all
other fields to zero and using for convenience the so--called
$\sigma$--parametrization for $U$ in the $N_f=2$ subspace:
\beq
{\cal L}_0^\gamma(1)({\rm pions})=e^2 C \langle Q U Q U^\dg\rangle
({\rm pions})= - \dfrac{2 e^2 C}{F^2}\pi^+\pi^-~,
\eeq 
where $Q$ is the quark charge matrix (\ref{eq:Qquark}) and $C$ is the 
unique LEC of $O(e^2 p^0)$. The important conclusion is that 
there are no terms of $O(\pi^n)$ for $n>2$. In
other words, the Lagrangian ${\cal L}_0^\gamma(1)$ contributes only
to the $\pi^+ - \pi^0$ mass difference, but not to the scattering
amplitude itself. To leading order, $O(e^2 p^0)$, electromagnetic 
corrections appear only in the kinematics and can easily be accounted for. 
The leading non--trivial electromagnetic effects for $\pi\pi$ scattering 
occur first at $O(e^2 p^2)$ and are expected to be quite a bit smaller
than suggested by Eq.~(\ref{eq:a00+0}).


\setcounter{equation}{0}
\setcounter{subsection}{0}
 
\section{Pions and Nucleons}
\label{sec:baryons}
In the meson--baryon system, we are still far from the precision attained
in the meson sector. There are several reasons for this difference:
\bit
\item
The baryons are not Goldstone fields and their interactions are therefore
less constrained by chiral symmetry (they are spectators, not
actors \cite{Leu94b}). 
\item
The chiral expansion for baryons has terms of all positive orders whereas 
only even orders appear in the meson case.
\item
Resonances are often closer to the physical thresholds [e.g., the 
$\Delta(1232)$ compared to the $\rho$ meson].
\eit
Nevertheless, there has been considerable progress in this field
both on the more theoretical and on the phenomenological side as well
(for recent reviews, cf. Refs.~\cite{BKM95,PMR93}). Chiral symmetry
and effective field theories are becoming the common basis for nuclear
and low--energy particle physics.

I will concentrate in this lecture on the pion--nucleon system
where most of the detailed work has been done up
to now. In comparison, for chiral $SU(3)$ mainly the non--analytic parts
of the chiral one--loop corrections have been calculated, but in most
cases there has not been a systematic treatment of the relevant LECs.

The question of LECs is especially acute in the meson--baryon system
because there are many of them and because they are sensitive 
to nearby strongly coupled baryon resonances. A special role
is played by the $\Delta(1232)$ or the decuplet in general. In the
framework of heavy baryon CHPT, the relevant quantity is the mass 
difference (usually also called $\Delta$)
\beq
\Delta := m_{\Delta} - m_N \simeq 3 F_\pi \simeq 2 M_\pi
\eeq
and the ratio $M_\pi/\Delta$. This ratio takes on three different values
for the limits of interest:
\beq
\dfrac{M_\pi}{\Delta} = \left\{\ba{ll}
0 & \qquad {\rm chiral \,\, limit} \\
0.5 & \qquad {\rm real \,\, world} \\
\infty & \qquad N_c \to \infty
\ea \right.~.
\eeq

The orthodox point of view is not to ascribe any special role to the 
baryon resonances (cf., e.g., Ref.~\cite{BKM95}). Like for the meson 
resonances, their influence is
contained in LECs some of which turn out to be rather big because the
$\Delta(1232)$ is not only close by but also strongly coupled to the
$\pi N$ system. The major advantage of the orthodox approach is that there 
is certainly no danger of double counting. On the other hand, for chiral 
$SU(3)$ one expands also in the strange quark mass. Since $M_K > \Delta$,
the standard chiral expansion will have very big LECs and the chiral
counting may become misleading even near threshold.

The other alternative is to include the decuplet as dynamical
degrees of freedom in the effective field theory \cite{JM91}.
In this effective Lagrangian, the LECs do not receive any
contributions from the decuplet and are therefore expected to be
much smaller. A systematic procedure to include the $\Delta(1232)$ not
only at leading order, but in principle to any order in heavy baryon
CHPT, has been presented only recently \cite{HHK96}. The idea is to expand
in $M_\pi$ (the meson masses in general) and in $\Delta$ at the
same time keeping the ratio $M_\pi/\Delta$ fixed at its physical value.
Although, as always for strongly decaying states, the problem of double 
counting is lurking in the back, this approach looks very promising
from both a theoretical and phenomenological point of view. 

In this last lecture, I will describe the orthodox approach of
heavy baryon CHPT for nucleons (and pions) only.

\subsection{From the linear $\sigma$ model to CHPT}
\label{subsec:sigma}
We are searching for the effective field theory describing the strong
interactions of pions and nucleons at low energies. As I emphasized
in the first lecture, this will be an effective field theory of the
non--decoupling type induced by spontaneous chiral symmetry breaking.
Such theories are generically nonrenormalizable. It does not make sense
in general to split the effective Lagrangian into renormalizable (marginal 
and relevant operators) and nonrenormalizable parts (irrelevant operators).

It is therefore remarkable that there is a renormalizable model
of pions and nucleons with spontaneous chiral symmetry breaking,
the linear $\sigma$ model \cite{SGML57}. Before discussing what 
price one has to pay for this non--generic feature of the model,
I will use it as an explicit example for making the structure of spontaneous
symmetry breaking transparent. By rewriting it in the form of a 
non--decoupling effective field theory, the abstract quantities 
introduced in the first lecture will emerge naturally.

We first rewrite the $\sigma$--model Lagrangian for the pion--nucleon system
\beq
\cL_\sigma = {1\over 2} \left(\del_\mu\sigma\del^\mu\sigma +
 \del_\mu\vec{\pi}\del^\mu\vec{\pi}\right) -{\lambda\over 4}
\left(\sigma^2 + \vec{\pi}^2 - v^2\right)^2 + \ol\psi
 ~i\not\!\partial\psi - g \ol\psi \left(\sigma + i\vec{\tau}\vec{\pi}
\gamma_5\right)\psi \label{eq:Lsig1}
\eeq
$$
\psi = { p \choose n}
$$
in the form
\beq
\cL_\sigma = {1\over 4}\langle \del_\mu\Sigma\del^\mu\Sigma \rangle
- {\lambda\over 16} \left( \langle \Sigma^\dg\Sigma \rangle - 2 v^2\right)
^2 + \ol{\psi_L} ~i\not\!\partial\psi_L + \ol{\psi_R} ~i\not\!\partial\psi_R
- g \ol{\psi_R} \Sigma\psi_L - g \ol{\psi_L} \Sigma^\dg\psi_R
\label{eq:Lsig2} \eeq
with
$$
\Sigma = \sigma{\bf 1} - i\vec{\tau}\vec{\pi} 
$$
to exhibit the chiral symmetry $G=SU(2)_L\times SU(2)_R$~:
$$
\psi_A \toG g_A \psi_A~,\qquad g_A \in SU(2)_A \quad(A=L,R)~,
\qquad \Sigma \toG g_R\Sigma g_L^{-1}~.
$$
For $v^2>0$, chiral symmetry is spontaneously broken and the
``physical" fields are the massive field $\hat\sigma = \sigma - v$
and the Goldstone fields $\vec \pi$. 

The Lagrangian (\ref{eq:Lsig1}) with its non--derivative couplings for 
the fields $\vec\pi$ immediately raises a question: how can these couplings
satisfy the Goldstone theorem that predicts a vanishing amplitude
for vanishing momenta of the Goldstone bosons? To answer
this question, we perform a field transformation from the original fields
$\psi$, $\sigma$, $\vec\pi$ to a new set $\Psi$, $S$, $\vec\vp$
through a polar decomposition of the matrix field $\Sigma$:
\beq
\Sigma = (v+S)U(\vp) ~,\quad\qquad \Psi_L=u\psi_L ~,\qquad
\Psi_R=u^\dg \psi_R \label{eq:ftrans}
\eeq
$$
S^\dg = S ~,\qquad U^\dg=U^{-1} ~,\qquad \det U=1 ~,\qquad U=u^2~.
$$
The matrix field $u(\vp)$ is precisely the coset element introduced
in (\ref{eq:coset}). The chiral transformation (\ref{eq:uphi}) of
$u(\vp)$ implies
\beq
U\toG g_R U g_L^{-1}~, \qquad S\toG S~, \qquad \Psi_A\toG h(g,\vp)\Psi_A
\quad (A=L,R)~.
\eeq
In the new fields, the $\sigma$--model Lagrangian (\ref{eq:Lsig1})
takes the form
\beqa
\cL &=& {v^2\over 4}(1 + {S\over v})^2\langle u_\mu u^\mu \rangle\no\\
&+& \ol\Psi ~i \not\!\nabla \Psi + {1\over 2} \ol\Psi
\not\!u \gamma_5\Psi - g(v+S)\ol\Psi \Psi + \ldots \label{eq:Lsig3}
\eeqa
with a covariant derivative $\nabla = \del + \Gamma$ and
\beqa
u_\mu(\vp) &=& i (u^\dg \partial_\mu u - u \partial_\mu u^\dg)
= i u^\dg \partial_\mu U u^\dg \no\\
\Gamma_\mu(\vp) &=& \frac{1}{2} (u^\dg \partial_\mu u
+ u \partial_\mu u^\dg) ~.
\label{eq:Gu}
\eeqa
The Lagrangian (\ref{eq:Lsig3}) allows for a clear separation
between the model independent Goldstone boson interactions
induced by spontaneous chiral symmetry breaking and the model dependent
part involving the scalar field $S$ [the kinetic term and the 
self--couplings of the scalar field are omitted in (\ref{eq:Lsig3})].
The first line in (\ref{eq:Lsig3}) is nothing but the nonlinear 
$\sigma$--model Lagrangian (\ref{eq:L2}), setting $S$ and all external
fields to zero ($v=F$).

The following conclusions emerge:
\bdes
\item[i.]
The $\sigma$--model Lagrangian in the form (\ref{eq:Lsig3}) has only
derivative couplings for the Goldstone bosons $\vec\vp$ contained in the 
matrix fields $u_\mu(\vp)$, $\Gamma_\mu(\vp)$. This answers the previous
question: the Goldstone theorem is now manifest at the Lagrangian level.
\item[ii.]
S--matrix elements are unchanged under the field
transformation (\ref{eq:ftrans}), but the Green functions are very
different. For instance, in the pseudoscalar meson sector the field $S$
does not contribute at all
at lowest order, $O(p^2)$, whereas $\hat\sigma$ exchange is essential
to repair the damage done by the non--derivative couplings of the
$\vec\pi$. A good exercise is to calculate the lowest--order $\pi\pi$ 
scattering amplitude in the two representations.
\item[iii.]
The manifest renormalizability of the Lagrangian (\ref{eq:Lsig1}) has
been traded for the manifest chiral structure of (\ref{eq:Lsig3}).
Of course, the Lagrangian (\ref{eq:Lsig3}) is still renormalizable,
but this renormali\-zability has its price. It requires specific
relations between various couplings that have nothing to do with
chiral symmetry and, which is worse, are not in agreement with
experiment. For instance, the model contains the Goldberger--Treiman
relation \cite{GT58} in the form ($m$ is the nucleon mass in the
chiral limit)
\beq
m = g v \equiv g_{\pi NN} F_\pi~.
\eeq
Instead of the physical value $g_A=1.26$ for the axial--vector
coupling constant $g_A$ the model has $g_A=1$. This by itself is not
enough to dismiss the model because higher--order corrections will
modify $g_A$. But it is one indication for the central problem of the
linear $\sigma$ model: although it has the right symmetries by construction, 
it is not general enough to describe the real world. The phenomenological
problems of the model are especially evident in the meson sector 
\cite{GL84}. The linear $\sigma$ model is an ingenious and illustrative
toy model, but it should not be mistaken for the effective field theory of 
QCD at low energies. The root of the problem is the requirement of
renormalizability.
\end{description}

Let me add one further remark to the perennial discussion about the
viability of the linear $\sigma$ model. Its fate as a realistic effective
field theory of QCD has almost nothing to do with the question
if and where a state with the quantum numbers of the $\sigma$ field
exists. There can hardly be any doubt that there is a 
$q \ol{q}$ state in the scalar spectrum with the same
quantum numbers although it is difficult
to isolate because of its huge width \cite{RT96}. However, the only
connection with chiral symmetry breaking is that the $\sigma$ field
has the right quantum numbers to mimic the quark condensates 
(\ref{eq:cond}). As a member of the scalar nonet, the corresponding
resonance plays its role in understanding some of the LECs of $O(p^4)$
\cite{EGPR89}. As far as we now know, it plays no role whatsoever for
spontaneous chiral symmetry breaking.

\subsection{Heavy nucleon expansion}
\label{subsec:HBCHPT}
Now that we have convinced ourselves that the linear $\sigma$ model
does not qualify as the effective field theory of QCD at low energies,
we have to construct the effective chiral Lagrangian in a more
systematic way. More specifically, we are after a systematic low--energy 
expansion of the pion--nucleon
Lagrangian for single--nucleon processes, i.e. for processes of the type
$\pi N \ra \pi \ldots \pi N$, $\gamma N \ra \pi \ldots \pi N$, 
$\gamma^* N \ra \pi 
\ldots \pi N$ (including nucleon form factors), $W^* N \ra \pi \ldots \pi N$.

There is an obvious problem with chiral counting here: in contrast to
pseudoscalar mesons, the nucleon four--momenta can never be ``soft" because
the nucleon mass does not vanish in the chiral limit. Although the problem
can be handled at the Lagrangian level \cite{GSS88,Krause90}, it reappears 
once one goes beyond tree level. The loop 
expansion and the derivative expansion do not coincide any more
like in the meson sector. The culprit is again the nucleon mass that enters 
loop amplitudes through the nucleon propagators. In the original relativistic
formulation \cite{GSS88}, amplitudes of a given chiral order
receive contributions from any number of loops.

A comparison between the nucleon mass and the chiral expansion scale
$4 \pi F_\pi$ suggests a simultaneous expansion in
$$
\dfrac{\vec{p}}{4\pi F} \qquad \mbox{and} \qquad \dfrac{\vec{p}}{m}
$$
where $\vec{p}$ is a small three--momentum.
On the other hand, there is a crucial difference between $F\simeq F_\pi$
and $m\simeq m_N$: whereas $F$ appears only in vertices, the
nucleon mass enters via the nucleon propagator. The
idea of heavy baryon CHPT \cite{JM91} is precisely to transfer 
$m$ from the propagators to some vertices.
The method can be interpreted as a clever choice of fermionic variables 
\cite{MRR92} in the generating functional of Green functions
\beq
e^{iZ[j,\eta,\bar\eta]} =  \int [du d\Psi d \bar \Psi]
\exp [i\{ S_M + S_{\pi N} + \int d^4 x (\bar \eta \Psi + \bar \Psi \eta)\}]~.
\label{eq:ZpiN}
\eeq
Here, $S_M + S_{\pi N}$ is the combined pion--nucleon action in the
relativistic framework, $\eta$ is a fermionic source and $j$ stands for 
the external fields $v,a,s,p$. 

In terms of velocity dependent fields $N_v,H_v$ defined as \cite{Geo90}
\beqa
N_v(x) &=& \exp[i m v \cdot x] P_v^+ \Psi(x) \label{eq:vdf} \\
H_v(x) &=& \exp[i m v \cdot x] P_v^- \Psi(x) \no \\
P_v^\pm &=& \frac{1}{2} (1 \pm \not\!v)~, \qquad v^2 = 1 ~, \no
\eeqa
with a time--like unit four--vector $v$, the pion--nucleon action 
$S_{\pi N}$ takes the general form
\beqa
S_{\pi N} &=& \int d^4 x \{ \bar N_v A N_v + \bar H_v B N_v +
\bar N_v \gamma^0 B^\dg \gamma^0 H_v - \bar H_v C H_v\} \label{eq:SpiN} \\
I &=& I_{(1)}+ I_{(2)} + I_{(3)} + \ldots~, \qquad I=A,B,C ~. \no 
\eeqa
The operators $A_{(n)}$, $B_{(n)}$, $C_{(n)}$ are the corresponding projections
of the relativistic pion--nucleon Lagrangians $\cL_{\pi N}^{(n)}$ 
\cite{GSS88,Krause90}. Rewriting also the source term in (\ref{eq:ZpiN}) 
in terms of $N_v, H_v$,
one can integrate out the ``heavy'' components $H_v$
 to obtain a nonlocal action in the 
fields $N_v$ \cite{BKKM92,EckPrag94}. Expanding this 
nonlocal action in a power series
in $1/m$, one obtains a Lorentz--covariant chiral expansion 
for the Lagrangian
\beq
\wh \cL_{\pi N}(N_v;v)=\wh \cL_{\pi N}^{(1)}+\wh \cL_{\pi N}^{(2)}+
\wh \cL_{\pi N}^{(3)}+\dots~, \label{eq:pinexp}
\eeq
which depends of course on the arbitrary
four--vector $v$. Specializing to the nucleon rest frame
with $v=(1,0,0,0)$, (\ref{eq:pinexp}) amounts to a 
nonrelativistic expansion for the $\pi N$ Lagrangian. In this Lagrangian,
the nucleon mass $m$ or rather its inverse appears only in vertices, but 
not in the propagator of the transformed nucleon field $N_v$.

A given Lorentz covariant Lagrangian for the field $N_v$
will in general not be Lorentz invariant because it depends on the
arbitrary four--vector $v$. To guarantee Lorentz invariance, two 
procedures are possible. One can either require that the Lagrangian
is unchanged under a change of $v$. This reparametrization 
invariance \cite{LM92} imposes relations on the couplings of the
Lagrangian (\ref{eq:pinexp}). Alternatively,
one can start directly from the fully relativistic Lagrangian which is 
Lorentz invariant by construction. The second approach has 
advantages especially in higher orders of the chiral expansion and it
implies of course reparametrization invariance. This is the
approach I am going to follow here.

The relativistic pion--nucleon Lagrangian of lowest order \cite{GSS88},
\beq
\cL_{\pi N}^{(1)} = \bar \Psi \left(i \not\!\nabla - m + \frac{g_A}{2}
\not\!u \gamma_5\right)\Psi, \label{eq:LMB1}
\eeq
leads directly to the corresponding ``nonrelativistic" Lagrangian of
$O(p)$:
\beq
\wh \cL_{\pi N}^{(1)} = \bar N_v(iv \cdot \nabla + g_A S \cdot u) N_v~.
\label{eq:piN1}
\eeq
Here,  $m$ and $g_A$ are the two LECs at $O(p)$, the nucleon mass and 
the axial--vector coupling
constant in the chiral limit. The covariant derivative $\nabla^\mu$ and
the vielbein field $u^\mu$ are defined in (\ref{eq:Gu}) except that
they now include external gauge fields like the photon. 
The spin matrix $S^\mu = i \gamma_5 \sigma^{\mu\nu} v_\nu/2$ is the only 
remnant of Dirac matrices in the Lagrangian $\wh \cL_{\pi N}$. By
construction, the mass $m$ has disappeared in the Lagrangian (\ref{eq:piN1})
and, consequently, the propagator of $N_v$ is independent of the mass. Note
that the relativistic Lagrangian (\ref{eq:LMB1}) has the same form as
the linear $\sigma$--model Lagrangian (\ref{eq:Lsig3}) except that
$g_A$ is arbitrary. 

At the next chiral order, $O(p^2)$, the Lagrangian $\wh \cL_{\pi N}^{(2)}$
consists of two pieces \cite{BKKM92}. There is first a piece 
that is due to the 
expansion in $1/m$ with completely determined coefficients and there
is in addition the nonrelativistic reduction of the relativistic
Lagrangian $\cL_{\pi N}^{(2)}$. After a suitable field transformation
of the nucleon field $N_v$, the
Lagrangian assumes its most compact form \cite{EM96}
\beqa
\wh \cL_{\pi N}^{(2)} &=& \bar N_v\left( - \frac{1}{2m} (\nabla \cdot 
\nabla + ig_A \{S \cdot \nabla, v \cdot u\}) \right. \no \\
&& \mbox{} + \frac{a_1}{m} \langle u \cdot u\rangle + 
\frac{a_2}{m} \langle (v \cdot u)^2\rangle +
\frac{a_3}{m} \langle \chi_+\rangle + 
\frac{a_4}{m} \left( \chi_+ - \frac{1}{2} \langle \chi_+\rangle \right)
\no \\
&& \left. \mbox{} + \frac{1}{m} \ve^{\mu\nu\rho\sigma} v_\rho S_\sigma 
[i a_5 u_\mu u_\nu + a_6 f_{+\mu\nu} + a_7 v_{\mu\nu}^{(s)}]\right) N_v~,
\label{eq:piN2}
\eeqa
where $\chi_+$ is related to $\chi$ defined in (\ref{eq:L2}) and thus
contains the quark masses. The tensor fields $f_{+\mu\nu}$, 
$v_{\mu\nu}^{(s)}$ are the isovector and isoscalar parts of the external 
gauge fields including the electromagnetic field.
The LECs $a_i$ ($i=1,\dots,7$) are dimensionless and expected to be
of $O(1)$ according to naive chiral dimensional analysis. Not surprisingly, 
some of them are bigger than the naive estimate because they account in 
particular for the effects of $\Delta(1232)$ exchange. The $a_i$ are
in fact all known phenomenologically and I refer to Bernard et al. 
\cite{BKM95} for an up--to--date review.  For future purposes, let me single 
out two of them that are related to the nucleon magnetic moments in the
chiral limit:
\beqa
\label{eq:nmm}
a_6 &=& \frac{\mu_v}{4} = \frac{1}{4} (\mu_p - \mu_n) \no \\
a_7 &=& \frac{\mu_s}{2} = \frac{1}{2} (\mu_p + \mu_n)~.
\eeqa

The first two terms in the Lagrangian (\ref{eq:piN2}) illustrate
the difference between Lorentz covariance and invariance. The latter
fixes the coefficients uniquely although covariance alone would
seem to allow arbitrary coefficients. That these coefficients
cannot be arbitrary becomes obvious when one realizes that the first
term governs the Thomson limit for nucleon Compton scattering
and that the
second one is responsible for the $O(p^2)$ contribution for pion
photoproduction on nucleons at threshold. Of course, both amplitudes are
completely determined by the nucleon charge and by $g_A$.

\subsection{The $\pi N$ Lagrangian to $O(p^3)$}
\label{subsec:piN}
At $O(p^3)$, loops enter the game. This can be seen most easily
from the formula for the chiral dimension $D$ [compare the corresponding
formula (\ref{eq:DL}) in the meson sector] of a general single--nucleon
diagram with $N_d^M$ mesonic vertices and $N_d^{MB}$ pion--nucleon vertices 
of $O(p^d)$ \cite{Wein90},
\beq
D = 2L + 1 + \sum_{d\ge 4} (d-2) N_d^M + \sum_{d\ge 2} (d-1) 
N_d^{MB} \geq 2L+1~. \label{eq:DMB2}
\eeq
The loop diagrams are in general divergent and must be 
regularized. As emphasized in the first lecture, the theory must therefore 
also be renormalized in order that the results are independent of the
regularization procedure. The divergences can be given in closed form 
as a Lagrangian with divergent coefficients \cite{Eck94}. As in the
meson sector, this Lagrangian must be a special case of the general
chiral Lagrangian of $O(p^3)$. To construct this Lagrangian is a little
more tedious than in the meson sector because all the relativistic
$\pi N$ Lagrangians of $O(p^n)$ with $n\le 3$ contribute to the
Lagrangian of $O(p^3)$ in heavy baryon CHPT. After applying
again suitable field transformations to get rid of redundant 
equation--of--motion terms, the complete $\pi N$ Lagrangian     
of $O(p^3)$ is found to be \cite{EM96}
\beqa
\label{eq:piN3}
\wh \cL_{\pi N}^{(3)} &=& \bar N_v \left( \frac{g_A}{8 m^2} 
[ \nabla_\mu,[\nabla^\mu,S \cdot u]] + \frac{1}{2m^2} \left[
\left\{ i \left( a_5 - \frac{1-3g^2_A}{8}\right) u_\mu u_\nu
 \right. \right. \right. \no \\
&& \mbox{} + \left. \left( a_6 - \frac{1}{8}\right) f_{+\mu\nu} +
\left( a_7 - \frac{1}{4}\right) v_{\mu\nu}^{(s)} \right\}
\ve^{\mu\nu\rho\sigma} S_\sigma i \nabla_\rho + \frac{g_A}{2}
S \cdot \nabla u \cdot \nabla \no \\
&& \mbox{} - \left. \frac{g^2_A}{8} \{v \cdot u,u_\mu\}
\ve^{\mu\nu\rho\sigma} v_\rho S_\sigma \nabla_\nu - \frac{ig_A}{16}
\ve^{\mu\nu\rho\sigma} f_{-\mu\nu} v_\rho \nabla_\sigma + {\rm h.c.}\right]
\no \\
&& \mbox{} + \left.\frac{1}{(4\pi F)^2} \sum_{i=1}^{24} b_i O_i \right)
N_v~.
\eeqa

Although quite a bit more involved, this Lagrangian has the same structure
as (\ref{eq:piN2}). There is a first part with coefficients completely fixed
in terms of LECs of $O(p)$ and $O(p^2)$. The second part has 24 new
LECs $b_i$. The associated field monomials $O_i$ can be
found in Ref.~\cite{EM96}. It is this second part that
is needed to absorb the divergences of the one--loop functional. The
splitting of the $b_i$ into divergent and finite parts introduces
again a scale dependence of the finite, measurable LECs $b_i^r(\mu)$. This
scale dependence is governed by $\beta$--functions that are determined
by the divergence functional \cite{Eck94}. Adding the finite part of
the one--loop functional, one arrives at the total generating
functional of Green functions in the pion--nucleon system up to $O(p^3)$:
\beq
Z = Z_1(g_A) + Z_2(a_i,g_A,m) +
 Z_{3,{\rm finite}}^{L=1}(g_A;\mu) + Z_3^{\rm tree}(b_i^r(\mu),
a_i,g_A,m)~. \label{eq:Z}
\eeq
The functionals $Z_1$, $Z_2$ are tree--level functionals, whereas
the functional of $O(p^3)$ consists of both a loop and a tree--level
part. The sum of these two and therefore the complete functional is
independent of the arbitrary scale $\mu$. Except for the complications
of heavy baryon CHPT, the situation is exactly the same as in the meson sector
to $O(p^4)$. By the way, if we want to extend the precision to
$O(p^4)$ also in the $\pi N$ sector, formula (\ref{eq:DMB2}) tells us
that we need again tree--level and one--loop diagrams only. In contrast
to $O(p^3)$, one--loop diagrams with a single vertex of $O(p^2)$ from 
(\ref{eq:piN2}) now also contribute.

The functional (\ref{eq:Z}) contains the complete low--energy structure
of the $\pi N$ system to $O(p^3)$. In order to extract physical amplitudes
from this functional, one has to calculate the appropriate
one--loop amplitudes contained in $Z_3^{L=1}$. This has already been
done for many processes of interest and I refer especially to 
Ref.~\cite{BKM95} for an extensive coverage of the 
phenomenological applications. 

Since we do not know very much yet about
the LECs $b_i$ of $O(p^3)$, those transitions that are insensitive to
the $b_i$ are especially interesting for comparison with experiment.
Those transitions can be divided into two groups. In
the first class, loop amplitudes do contribute, which are then necessarily
finite because there are by definition no counterterms that
could absorb the divergences. In the second type of amplitudes,
there are neither loop nor counterterm contributions at $O(p^3)$. The only 
possible other contribution at this order must then be due to the terms 
with fixed coefficients in the first part of (\ref{eq:piN3}).
In the last part of this lecture, I am going to discuss one example of each
class, both of them occurring in nucleon Compton scattering.

\subsection{Nucleon Compton scattering}
\label{subsec:Compton}
Nucleon Compton scattering
\beq
\gamma(k) + N(p) \to \gamma(k') + N(p')
\eeq
is described by six invariant amplitudes. The complete calculation to
$O(p^3)$ can be found in Ref.~\cite{BKKM92}. Here, I will limit
myself to the forward direction ($k=k'$, $p=p'$). 
In a gauge where the
polarization vectors have vanishing time components, the forward 
scattering amplitude can be written in the form 
\beq
T = f_1(\omega) \vec{\ve}\,' \cdot \vec{\ve} + i  \omega 
f_2(\omega) \vec{\sigma} \cdot (
\vec{\ve}\,' \times \vec{\ve} ) ~ ,
\label{eq:compton}
\eeq
where $\omega=k \cdot v$ is the photon energy in the nucleon rest
frame. 

Let us first take a look at the spin--flip amplitude $f_2(\omega)$.
Since an external photon counts like a momentum for the chiral dimension,
the leading contribution to $f_2$ appears at $O(p^3)$ because of the
explicit factor $\omega$ in front of $f_2$ in (\ref{eq:compton}).
It is straightforward to check that in the limit of vanishing
photon energy there are neither loop nor counterterm contributions with
the LECs $b_i$. Therefore, $f_2(0)$ must be expressible in
terms of $g_A$ and the LECs $a_i$ of $O(p^2)$ only, according to 
(\ref{eq:piN3}) and (\ref{eq:Z}). 

\begin{figure}
\centerline{\epsfig{file=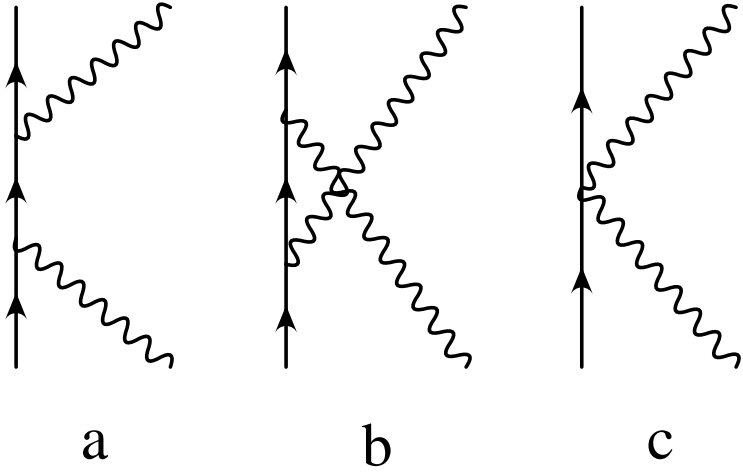,height=7cm}}
\caption{Tree diagrams for nucleon Compton scattering.}
\label{fig:compton}
\end{figure}

How does heavy baryon CHPT account for the leading spin--dependent Compton 
amplitude $f_2(0)$? The relevant diagrams
are shown in Fig. \ref{fig:compton} where the vertices in diagrams a,b
are due to the Lagrangian (\ref{eq:piN2}), while the seagull vertex
of diagram c comes from the $O(p^3)$ Lagrangian (\ref{eq:piN3}).
From these Lagrangians one extracts the respective couplings (up to
trivial factors)
\beqa
k_2 &=& a_6 \tau_3 + \frac{a_7}{2} = \frac{1}{2}
\left( \ba{cc} 1 + \kappa_p & 0 \\ 0 & \kappa_n \ea \right) \\*
k_3 &=& \frac{1}{2} (1+ \tau_3) \left[ \left(a_6 - \frac{1}{8}\right)
\tau_3 + \frac{1}{2} \left( a_7 - \frac{1}{4}\right) \right] =
\frac{1}{4} \left( \ba{cc} 1 + 2\kappa_p & 0 \\ 0 & 0 \ea \right)\no
\eeqa
in terms of the nucleon anomalous magnetic moments $\kappa_N$
[cf. Eq.~(\ref{eq:nmm})]. 

It is clear that the sum of diagrams a,b is proportional to
$k_2^2$ while the seagull diagram c is proportional to $k_3$. A simple
calculation shows that the sum of all three diagrams is
governed by
\beq
k^2_2 - k_3 = \frac{1}{4} \left( \ba{cc} \kappa^2_p & 0 \\ 0 & \kappa^2_n
\ea \right)~.
\eeq
Putting in the appropriate factors, one recovers the classic low--energy 
theorem \cite{LGMG54}
\beq
f_2(0) = - \dfrac{e^2 \kappa_N^2}{2 m^2}~.
\eeq

Turning now to the spin independent amplitude, let me write it
$f_1(\omega,M_\pi)$ in order to make the dependence on quantities with
non--vanishing chiral dimension manifest. As already discussed
in the second lecture, the chiral dimension of an amplitude has to
do with the degree of homogeneity of the amplitude in momenta and masses.
Except for chiral logs which have an additional dependence
on $M_\pi/\mu$, the number $D_L$ defined in (\ref{eq:DF}) is the degree 
of homogeneity of the function
$f_1(\omega,M_\pi)$ in the chiral variables $\omega$, $M_\pi$.
With $D_F = 2$ external photons, $D_L$ 
is obtained from (\ref{eq:DMB2}) as
\beq
D_L = 2L - 1 + \sum_{d\ge 4} (d-2)N_d^M + \sum_{d\ge 2} (d-1)N_d^{MB} 
\geq - 1~.
\label{eq:D_L}
\eeq
Therefore, the chiral expansion of
$f_1(\omega,M_\pi)$ assumes the general form \cite{EM95}
\beq
f_1(\omega,M_\pi) = \frac{1}{\omega} g_{-1}(\omega/M_\pi) + g_0(\omega/M_\pi)
+ \sum_{D_L \geq 1} \omega^{D_L} g_{D_L}(\omega/M_\pi)
\label{eq:fomega} \eeq
with functions $g_i$ depending only on the ratio $\omega/M_\pi$.
Eq.~(\ref{eq:D_L}) shows that only tree diagrams can contribute to the first 
two terms. Since the relevant tree diagrams \footnote{The diagrams are again
the ones in Fig.~\ref{fig:compton} although the couplings are not
the same as for the spin--flip amplitude.} do not contain pion lines, 
$g_{-1}$ and $g_0$ must actually be constants. Explicit calculation 
\cite{BKKM92} reproduces the Thomson limit ($q_N$ is the nucleon charge 
in units of $e$)
\beq
g_{-1} = 0~, \qquad g_0 = f_1(0,M_\pi) = - \frac{e^2 q_N^2}{m}
\eeq
since \cite{EM95}
\beq
\lim_{\omega \ra 0} \omega^{n-1} g_n(\omega/M_\pi) = 0 \qquad
(n \geq 1)~.
\eeq

\begin{table}[t]
\caption{Comparison between theory \protect\cite{BKMS93} and
experiment \protect\cite{Nathan94} for the electromagnetic polarizabilities 
of the nucleons in units of $10^{-4}$~fm$^3$. The experimental errors
are anticorrelated because in the analysis a model independent dispersion sum
rule is used that determines
$\alpha + \beta$ with a relatively small error.}
\label{tab:polar}
$$
\begin{tabular}{|c|c|c|} \hline
polarizability & CHPT to $O(p^4)$ & data \\ \hline
$\alpha_p$ & $10.5 \pm 2.0$ & $\qquad 12.0 \pm 0.8 \pm 0.4\qquad $ \\
$\beta_p$  & $3.5 \pm 3.6$  & $2.2 \mp 0.8 \mp 0.4$ \\
$\alpha_n$ & $13.4 \pm 1.5$ & $12.5 \pm 1.5 \pm 2.0$ \\
$\beta_n$   & $7.8 \pm 3.6$  & $3.5 \mp 1.8 \mp 2.0$ \\ \hline
\end{tabular}
$$
\end{table}

The function $g_1(\omega/M_\pi)$ in (\ref{eq:fomega}) is a quantity of
$O(p^3)$ ($D_L=1$). In this case, there are no contributions at all
from the third--order Lagrangian (\ref{eq:piN3}) and so the loop amplitude
must be finite. From the one--loop result \cite{BKM91,BKKM92} I extract
the term linear in $\omega$,
\beq
g_1(\omega/M_\pi) = \frac{11 e^2 g_A^2 \omega}{192 \pi F^2 M_\pi}
+ O(\omega^2)~,
\eeq
because this linear term is proportional to the sum of the electric
and magnetic polarizabilities $\alpha + \beta$. A similar calculation
for the non--forward amplitude produces the difference $\alpha - \beta$.
Altogether, one finds for the polarizabilities to leading order, $O(p^3)$,
in the chiral expansion \cite{BKM91,BKKM92}
\beqa
\alpha_p = \alpha_n &=& \frac{5 e^2 g_A^2}{384 \pi^2 F_\pi^2
M_\pi} = 12 \cdot 10^{-4} \mbox{ fm}^3 \\
\beta_p =  \beta_n &=& \frac{\alpha_p}{10} =
1.2 \cdot 10^{-4} \mbox{ fm}^3~.
\eeqa

Comparison with the experimental values in Table \ref{tab:polar} shows
that the long--range pion cloud clearly accounts for the main features
of the data. This is as it should be because 
the representation (\ref{eq:fomega}) implies that the
contribution of $O(p^n)$ $(D_L = n-2)$ to the polarizabilities is of
the form $c_n M_\pi^{n-4}$ $(n \geq 3)$ with constant $c_n$. Higher--order
contributions to the polarizabilities are suppressed by additional factors
of $M_\pi$.

Instead of discussing in detail the results of a calculation to $O(p^4)$ 
\cite{BKMS93} listed in Table \ref{tab:polar}, let me emphasize once more the
importance of the leading--order result. It is certainly gratifying that
experiment is in excellent agreement with the CHPT prediction. But from a
theoretical point of view it is also important to realize that
CHPT has produced a general low--energy theorem: 
\beq
\lim_{M_\pi \to 0} M_\pi \alpha_N = 10\,\lim _{M_\pi \to 0} M_\pi \beta_N
 = \frac{5 e^2 g_A^2}{384 \pi^2 F_\pi^2}~.
\eeq
This is an unambiguous consequence of the Standard Model.
In other words, if a specific hadronic model for the nucleon polarizabilities
does not satisfy this low--energy theorem it may have all possible
phenomenological virtues but it can certainly not be in agreement
with QCD.

\vfill

\section*{Acknowledgements}
\noindent
I am very grateful to the organizers of this Workshop, in
particular to Erasmo Ferreira and Jos\'e de S\'a Borges,
for the opportunity to lecture in such beautiful surroundings  
and especially for the wonderful hospitality during my stay in Brazil.
This work has been supported in part by FWF (Austria), Project 
No. P09505--PHY
and by HCM, EEC--Contract No. CHRX--CT920026 (EURODA$\Phi$NE).

\newpage

\newcommand{\PL}[3]{{Phys. Lett.}        {#1} {(19#2)} {#3}}
\newcommand{\PRL}[3]{{Phys. Rev. Lett.} {#1} {(19#2)} {#3}}
\newcommand{\PR}[3]{{Phys. Rev.}        {#1} {(19#2)} {#3}}
\newcommand{\NP}[3]{{Nucl. Phys.}        {#1} {(19#2)} {#3}}

\end{document}